\documentclass[aps,preprintnumbers,twocolumn,amsmath,amssymb,nofootinbib]{revtex4}

\usepackage{amsmath,amssymb}
\usepackage{graphicx}
\newcommand{\amp}{&\!\!}
\newcommand{\beq}{\begin{equation}}
\newcommand{\eeq}{\end{equation}}
\newcommand{\bea}{\begin{eqnarray}}
\newcommand{\eea}{\end{eqnarray}}

\begin{document}

\author{
Mark P. Hertzberg\footnote[0]{Electronic address: mphertz@mit.edu}}
\address{Center for Theoretical Physics\\ Massachusetts Institute of Technology, Cambridge, MA 02139, USA}


\title{Quantum Radiation of Oscillons}

\begin{abstract}
Many classical scalar field theories possess remarkable solutions: coherently oscillating, localized clumps, known as oscillons. 
In many cases, the decay rate of classical small amplitude oscillons is known to be exponentially suppressed and so they are extremely long lived.
In this work we compute the decay rate of quantized oscillons. 
We find it to be a power law in the amplitude and couplings of the theory. Therefore, the quantum decay rate is very different to the classical decay rate and is often dominant.
We show that essentially all oscillons eventually
decay by producing outgoing radiation.
In single field theories the outgoing radiation has typically linear growth,
while if the oscillon is coupled to other bosons the outgoing radiation can have exponential growth.
The latter is a form of parametric resonance: explosive energy transfer from a localized clump into daughter fields.
This may lead to interesting phenomenology in the early universe. 
Our results are obtained from a perturbative analysis, a non-perturbative Floquet analysis, and numerics.

\end{abstract}
\vspace*{-\bigskipamount} \preprint{MIT-CTP-4130}

\maketitle

\section{Introduction}
Although there is no direct evidence yet, there are good reasons to think 
that scalar fields are plentiful in nature,
such as the Higgs boson, axion, inflaton, p/reheating fields, moduli, squarks, sleptons, etc. In recent years, there has been increasing interest in certain kinds of long lived structures that scalar fields can support. In particular, under fairly broad conditions, 
if a massive scalar field $\phi$ possesses a nonlinear self-interacting potential, such as $\phi^3+\ldots$ or $-\phi^4+\ldots$, then it can support coherently oscillating localized clumps, known as oscillons. Due to their oscillations in time and localization in space, they are time-dependent solitons.

Remarkably, oscillons have a long lifetime, i.e., they may live for very many oscillations, despite the absence of any internal conserved charges. Related objects are Q-balls \cite{Coleman}, which do carry a conserved charge. If the field $\phi$ is promoted to a complex scalar carrying a U(1) global symmetry 
$\phi\to e^{i\alpha}\phi$, then oscillons correspond to oscillatory radial motion in the complex $\phi$ plane, while Q-balls correspond to circular motion in the complex $\phi$-plane. Since oscillons do not require the U(1) symmetry for their existence (in fact they are typically comprised of only a real scalar, as we will assume), then oscillons appear even more generically than Q-balls.

As classical solutions of nonlinear equations of motion, oscillons (related to ``quasi-breathers") possess an asymptotic expansion which is exactly periodic in time and localized in space, characterized by a small parameter $\epsilon$.
In a seminal paper Segur and Kruskal \cite{Segur} showed that the asymptotic expansion is not an exact solution of the equations of motion and in fact misses an exponentially small radiating tail, which they computed. They showed that oscillons decay into outgoing radiation at an exponentially suppressed rate $\sim\exp(-b/\epsilon)$, where $b=\mathcal{O}(1)$ is model dependent. Although their work was in 1 spatial dimension, similar results have been obtained in 2 and 3 spatial dimensions \cite{Fodor:2009kf}. 
Furthermore, various long lived oscillons 
have been found in different contexts, including the standard model \cite{Farhi:2005rz,Graham:2006vy,Graham:2007ds}, abelian-Higgs models \cite{Gleiser:2008dt}, axion models \cite{Kolb}, in an expanding universe \cite{Graham:2006xs,Farhi:2007wj}, during phase transitions \cite{Copeland:1995fq,Dymnikova:2000dy,Gleiser:2006te,Gleiser:2007ts}, domain walls \cite{Hindmarsh:2007jb},
gravitational systems \cite{Alcubierre:2003sx,Page:2003rd,Fodor:2009kg} (called ``oscillatons"), large $+\phi^6$ models \cite{Mustafa}, 
in other dimensions \cite{Saffin:2006yk}, etc, showing that oscillons are generic and robust. 

In almost all cases, investigations into oscillons have thus far been at the classical level. 
There is a good reason for this approximation: the mass of an oscillon $M_{\mbox{\tiny{osc}}}$ is typically much greater than the mass of the individual particles $m_\phi$. This means that the spectrum is almost continuous and classical. But does this imply that every property of the oscillon, including its lifetime, is adequately described by the classical theory? 
To focus the discussion, suppose a scalar field $\phi$ in $d+1$ dimensional Minkowski space-time has a potential of the form
\begin{eqnarray}
V(\phi)=\frac{1}{2}m^2\phi^2 +{\lambda_3\over 3!}m^{(5-d)/2}\phi^3+\frac{\lambda_4}{4!}m^{3-d}\phi^4+\ldots
\end{eqnarray}
Since we will keep explicit track of $\hbar$ in this paper, it has to be understood that $m$ here has units $T^{-1}$; the quanta have mass $m_\phi=\hbar\,m$, and \{$\lambda_3^2,\,\lambda_4$\} have units $\hbar^{-1}$. It is known that for 
$\lambda\equiv{5\over3}\lambda_3^2-\lambda_4>0$ such a theory possesses classical oscillons that are characterized by a small dimensionless parameter $\epsilon$ and have a mass given by
$M_{\mbox{\tiny{osc}}}\sim\frac{m}{\lambda}l(\epsilon) = \frac{m_\phi}{\lambda\,\hbar}l(\epsilon)$,
where $l(\epsilon)=\epsilon^{2-d}$ in the standard expansion that we will describe in Section \ref{Classical}, but can scale differently in other models (e.g., see \cite{Fodor:2009kg}).
So if $\lambda\,\hbar\ll l(\epsilon)$, then $M_{\mbox{\tiny{osc}}}\gg m_\phi$. 

In such a regime we expect that various properties of the oscillon are well described classically, such as its size and shape. In this work we examine whether the same is true for the oscillon lifetime. We find that although the oscillon lifetime is exponentially long lived classically, it has a power law lifetime in the quantum theory controlled by the ``effective $\hbar$" -- in the above case this is $\lambda_3^2\,\hbar$ or $\lambda_4\,\hbar$.

In this paper we treat the oscillon as a classical space-time dependent background, as defined by the $\epsilon$ expansion, and quantize field/s in this background. We go to leading order in $\hbar$ in the quantum theory. 
We find that oscillon's decay through the emission of radiation with wavenumbers $k=\mathcal{O}(m)$. We show that the decay is a power law in $\epsilon$ and the couplings, and we explain why this is exponentially suppressed classically. 
Our analysis is done for both single field theories, where the emitted radiation typically grows at a linear rate corresponding to $3\,\phi\to2\,\phi$ or $4\,\phi\to2\,\phi$ annihilation processes, and for multi-field theories, where the emitted radiation often grows at an exponential rate corresponding to $\phi\to2\,\chi$ decay or $2\,\phi\to2\,\chi$ annihilation processess. 
We calculate the quantum decay rates in several models, which is supported by numerical investigations, 
but our work is also qualitative and of general validity. We also comment on collapse instabilities for $k=\mathcal{O}(\epsilon\,m)$ modes, whose existence is model dependent.

The outline of this paper is as follows: In Section \ref{Classical} we start with a review of classical oscillons and describe their 
exponentially suppressed decay in Section \ref{ExpSmall}.
In Section \ref{Quantum} we outline the semi-classical quantization of oscillons and derive the decay rate of oscillons in Section \ref{HigherQuantum}. Having started with single field models, we move on to examine the effects of coupling to other fields in Section \ref{TwoField}. 
In Section \ref{ExpLinear} we discuss when the decay products grow linearly in time and when it is exponential. Here we demonstrate that coupled fields can achieve (depending on parameters) explosive energy transfer, which may have some cosmological relevance. We comment on collapse instabilities in Section \ref{LeadingQuantum} and conclude in Section \ref{Conclusions}.

\section{Classical Oscillons}\label{Classical}

In this section we review the asymptotic expansion for a single field oscillon, valid in 1, 2, and 3 dimensions.
We then explain why the oscillon slowly radiates at an exponentially suppressed rate.


Consider a single scalar field $\phi$ in $d+1$ dimensional Minkowski space-time (signature $+ - - \ldots$) with classical action
\begin{eqnarray}
S=\int d^{d+1}x\left[\frac{1}{2}(\partial\phi)^2
-\frac{1}{2}m^2\phi^2 -V_I(\phi)\right],
\end{eqnarray}
where $V_I(\phi)$ is a nonlinear interaction potential. Here $m$ has units $T^{-1}$. For simplicity we measure time in units of $1/m$, so without loss of generality, we set $m=1$ from now on in the paper, unless otherwise stated. 
The classical equation of motion is
\begin{eqnarray}
\ddot{\phi}-\nabla^2\phi+\phi+V_I'(\phi)=0.
\label{eom}
\end{eqnarray}
In this paper we will focus on the following types of interaction potentials
\begin{eqnarray}
V_I(\phi) & = & \frac{\lambda_3}{3!}\phi^3+\frac{\lambda_4}{4!}\phi^4+\ldots,
\end{eqnarray}
where  $\lambda\equiv {5\over3}\lambda_3^2-\lambda_4>0$ is assumed.
This includes the cases (i) $\lambda_3> 0$ and $\lambda_4>0$, which will occur for a generic symmetry breaking potential,
and (ii) $\lambda_3=0$ and $\lambda_4<0$, which is relevant to examples such as the axion. In case (ii) higher order terms in the potential, such as $+\phi^6$ are needed for stabilization.
Note that in both cases the interaction term causes the total potential to be reduced from the pure $\phi^2$ parabola. This occurs for $\phi<0$ in the $\phi^3$ case, and for $|\phi|>0$ in the $\phi^4$ case.
It is straightforward to check that a ball rolling in such a potential will oscillate at a frequency {\em lower} than $m=1$. If the field $\phi$ has spatial structure, then one can imagine a situation in which the gradient term in eq.~(\ref{eom}) balances the nonlinear terms, and a localized structure oscillates at such a low frequency -- this is the oscillon. 

Naively, this low frequency oscillation cannot couple to normal dispersive modes in the system and should be stable. Higher harmonics are generated by the nonlinearity, but resonances can be cancelled order by order in a small amplitude expansion. Let us briefly explain how this works -- more detailed descriptions can be found in Refs.~\cite{Kichenassamy,Fodor:2008es}.

In order to find periodic and spatially localized solutions, it is useful to rescale time $t$ and lengths $x=|{\bf x}|$ to
\beq 
\tau=t\,\sqrt{1-\epsilon^2},\,\,\,\,\,\,\,\, \rho=x\,\epsilon, 
\eeq
with $0<\epsilon\ll 1$ a small dimensionless parameter. Here we search for spherically symmetric solutions. The equation of motion (\ref{eom})  becomes
\begin{eqnarray}
(1-\epsilon^2)\partial_{\tau\tau}\phi-\epsilon^2\!\left(\!\partial_{\rho\rho}\phi+{d-1\over\rho}\partial_\rho\phi\!\right)\!+\phi+V_I'(\phi)=0\,\,
\label{eom2}\end{eqnarray}
To obtain an oscillon solution, the field $\phi$ is expanded as an asymptotic series in powers of $\epsilon$ as
\begin{eqnarray}
\phi_{\mbox{\tiny{osc}}}(\rho,\tau)=\sum_{n=1}^\infty \epsilon^n\,\phi_n(\rho,\tau).
\end{eqnarray}
The set of $n$ depends on $V_I(\phi)$. For $V_I\sim\lambda_3\,\phi^3$; $n=1,2,3,\ldots$, and for $V_I\sim-|\lambda_4|\,\phi^4$; $n=1,3,5,\ldots$. Upon substitution of the series into eq.~(\ref{eom2}), the leading term must satisfy 
\begin{eqnarray}
\partial_{\tau\tau}\phi_1+\phi_1=0,
\end{eqnarray}
with solution $\phi_1=f(\rho)\cos\tau$, where $f(\rho)$ is some spatial profile. Since
$\tau=t\,\sqrt{1-\epsilon^2}$ the fundamental frequency of oscillation is evidently $\omega=\sqrt{1-\epsilon^2}<1$. 

\begin{figure}[t]
\center{\includegraphics[width=\columnwidth]{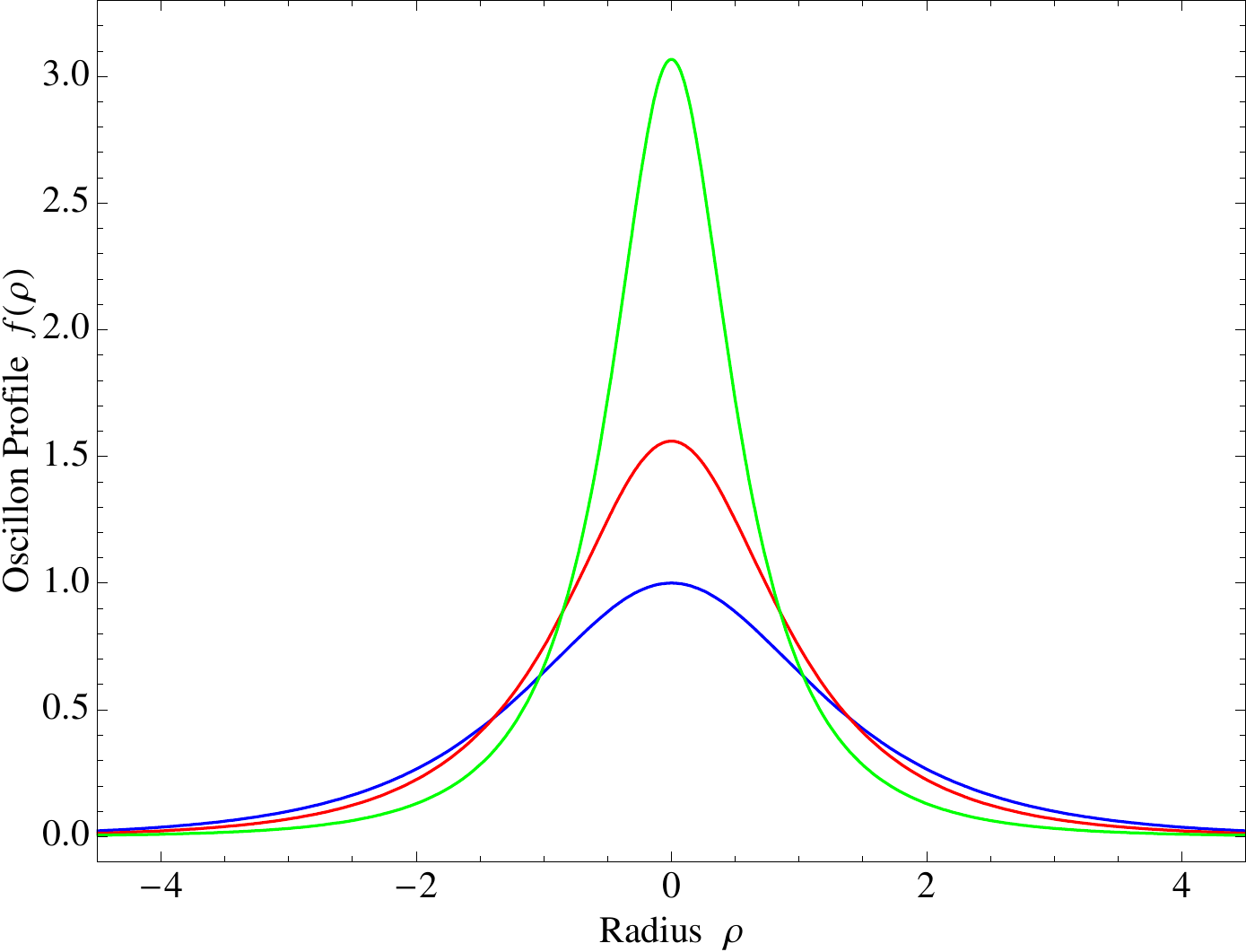}}
\caption{Leading order oscillon profile $\tilde{f}(\rho)$ in $d=1$ (blue), $d=2$ (red), and $d=3$ (green) dimensions. We allow $\rho$ to take on both positive and negative values here, i.e., a 1-d slice through the origin.}
\label{Profile}
\end{figure}

The next order terms in the expansion must not be resonantly driven by $\phi_1$, or the solution would not be periodic. By writing down the equations for $\phi_2$ and $\phi_3$ and demanding that the driving terms are non-resonant, we
establish an ODE for $f(\rho)$. Extracting the $\lambda$ dependence by defining $f(\rho)=4\tilde{f}(\rho)/\sqrt{\lambda}$, the ODE is found to be
\begin{eqnarray}
\partial_{\rho\rho}\tilde{f}+{d-1\over\rho}\partial_\rho\tilde{f}-\tilde{f}+2\tilde{f}^3=0.
\end{eqnarray}
This ODE possesses a localized solution for $d=1,2,3$.
In $d=1$ the solution is known analytically $\tilde{f}(\rho)=\mbox{sech}\,\rho$, but is only known numerically in $d=2$ and $d=3$. (For $d=2,3$ there are infinitely many solutions, of which we take the fundamental solution). We plot $\tilde{f}(\rho)$ in Figure \ref{Profile}.
Altogether this gives the leading order term for the oscillon
\begin{eqnarray}
\phi_{\mbox{\tiny{osc}}}(\rho,\tau)=\frac{4\,\epsilon}{\sqrt{\lambda}}\tilde{f}(\rho)\cos\tau
+\sum_{n>1}^\infty \epsilon^n\,\phi_n(\rho,\tau),
\label{phi1}
\end{eqnarray}
where higher order terms can be obtained in a similar manner. At any finite order in $\epsilon$ this provides a periodic and spatially localized oscillon.
By integrating the oscillon's energy density over all space, the total mass of the oscillon is found to scale to leading order in $\epsilon$ as $M_{\mbox{\tiny{osc}}}\sim m\,\epsilon^{2-d}/\lambda$. 

\section{Classical Radiation}\label{ExpSmall}

In 1987 Segur and Kruskal \cite{Segur} found that the above asymptotic expansion is not exact, as it misses an exponentially small radiating tail. They computed this using matched expansions between the inner core of the oscillon and infinity, involving some detailed analysis.
Here we describe the physical origin of this classical radiation in simple terms.

The existence of outgoing radiation is ultimately tied to the fact that the oscillon expansion in not an exact solution of the equations
of motion, it is only an asymptotic expansion which is correct order by order in $\epsilon$, but not beyond all orders. Consider the oscillon expansion, truncated to order $N$, i.e.,
\begin{equation}
 \phi_{\mbox{\tiny{osc}}}(x,t)=\sum_{n=1}^N \epsilon^n\,\phi_n(x,t).
\end{equation}
Lets substitute this back into the equation of motion. This will involve many harmonics. By construction the fundamental mode $\cos\tau=\cos(\omega\,t)$ will cancel, since the spatial structure was organized in order to avoid such a resonance.
However, we do obtain remainders including higher harmonics.  
This leads to the following remainder
\begin{equation}
J(x,t)=j(x)\cos(\bar{n}\,\omega\,t)+\ldots,
\end{equation}
where for $d=1$ we have
\beq
j(x)=C_N \epsilon^{N+2}\mbox{sech}^{N+2}(x\epsilon)+\ldots,
\eeq
where we have included only the coefficient of the next harmonic, with some coefficient $C_N$. 
$\bar{n}=2$ for asymmetric potentials and $\bar{n}=3$ for symmetric potentials.

Let us now decompose a solution of the equations of motion $\phi_{\mbox{\tiny{sol}}}$ into an oscillon piece and a 
correction $\delta$
\beq
\phi_{\mbox{\tiny{sol}}}(x,t)=\phi_{\mbox{\tiny{osc}}}(x,t)+\delta(x,t)
\eeq
with $\delta$ taking us from the oscillon expansion to the nearest solution.
We find that the remainder $J(x,t)$ acts as a source for the correction $\delta$
\begin{equation}
\ddot{\delta}-\nabla^2\delta+\delta=-J(x,t)
\label{classFO}\end{equation}
where we have ignored non-linear terms ($\delta^2$ etc) and a parametric driving term $V''(\phi_{\mbox{\tiny{osc}}})\delta$ (although such a term will be important in the quantum theory).
The solution in the far distance regime is obtained by standard means
\bea
\delta(x,t)&=& - \int {d^dk\,d\omega\over(2\pi)^{d+1}} \frac{J(k,\omega)e^{i({\bf k}\cdot{\bf x}-\omega t)}}{-\omega^2+k^2+1\pm i.0^+}\label{FTint}\\
&\sim& {\cos(k_{\mbox{\tiny{rad}}}x+\gamma)\over x^{(d-1)/2}}\cos(\omega_{\mbox{\tiny{rad}}} t)j(k_{\mbox{\tiny{rad}}}),
\eea
as $x\to\infty$ ($\gamma$ is a phase), suggesting that there is a radiation tail with an amplitude determined by the Fourier transform of the spatial structure of the source, evaluated at $k_{\mbox{\tiny{rad}}}$.  The kinematics of (\ref{FTint}) says that the radiation tail has frequency and wavenumber:
\begin{eqnarray}
w_{\mbox{\tiny{rad}}}=\bar{n}\,\omega,\,\,\,\,\,\,k_{\mbox{\tiny{rad}}}=\sqrt{\bar{n}^2\omega^2-1},
\label{kinematics}\end{eqnarray}
where $\omega=\sqrt{1-\epsilon^2}$ is the fundamental frequency of the oscillon.
This is the so-called ``quasi-breather" which is used to match onto a radiating oscillon \cite{Fodor:2009kf}.

Let us now evaluate $j(k)$. In $d=1$ we find
\begin{equation}
j(k)=C_N k^{N+1}\mbox{sech}\left(\frac{\pi\,k}{2\,\epsilon}\right)+\ldots
\end{equation}
So for $k=\mathcal{O}(\epsilon)$ the remainder is a power law in $\epsilon$, but for $k=\mathcal{O}(1)$, the prefactor is $\mathcal{O}(1)$ with a sech function evaluated in its tail. For $k=\mathcal{O}(1)$, which is the case for $k=k_{\mbox{\tiny{rad}}}$, we note that the source is comparable to the Fourier transform of the oscillon itself, i.e.,
\beq
j(k)\sim\phi_{\mbox{\tiny{osc}}}(k)
\eeq
Lets now compute the spatial Fourier transform of the oscillon. Taking only the leading order piece from eq.~(\ref{phi1}) at $t=0$
we have
\begin{eqnarray}
\phi_{\mbox{\tiny{osc}}}(k)= \frac{4\,\epsilon}{\sqrt{\lambda}}\int d^dx\, \tilde{f}(x\epsilon)e^{i{\bf k}\cdot{\bf x}}.
\label{FT2}\end{eqnarray}
For $d=1$ we have $\tilde{f}(\rho)=\mbox{sech}\,\rho$, allowing us to compute the Fourier transform analytically:
\begin{eqnarray}
\phi_{\mbox{\tiny{osc}}}(k)=\frac{4\pi}{\sqrt{\lambda}}\mbox{sech}\left(\frac{\pi\,k}{2\,\epsilon}\right).
\label{FTsech}\end{eqnarray}
For $k\gg\epsilon$ this is exponentially small. The same is true in $d=2$ and $d=3$ which can be computed numerically. We can summarize the behavior for $d=1,2,3$ for $k\gg\epsilon$ by the following scaling 
\begin{eqnarray}
\phi_{\mbox{\tiny{osc}}}(k)\sim\frac{1}{\sqrt{\lambda}\,(\epsilon\,k)^{(d-1)/2}}\exp\left(-\frac{c_d\,k}{\epsilon}\right),
\label{FTexp}\end{eqnarray}
where the constant in the argument of the exponential is dimensional dependent: $c_1=\pi/2$, $c_2\approx1.1$, and $c_3\approx0.6$.\footnote{The constant $c_d$ is in fact the first simple pole of $\tilde{f}(\rho)$ along the imaginary $\rho$-axis; see Ref.~\cite{Fodor:2009kf} for comparison.}
If we evaluate this at the relevant mode $k=k_{\mbox{\tiny{rad}}}\approx\sqrt{\bar{n}^2-1}$,
 the Fourier amplitude is exponentially small as $\epsilon\to 0$. 

Hence the longevity of the oscillon is due to the fact that the spatial structure is dominated by small $k=\mathcal{O}(\epsilon)$ wavenumbers, while any outgoing radiation occurs through $k=\mathcal{O}(1)$ wavenumbers, whose amplitude is suppressed.
 The rate at which energy is lost scales as 
 \begin{eqnarray}
\left|\frac{dE_{\mbox{\tiny{osc}}}}{dt}\right|  \sim |\phi_{\mbox{\tiny{osc}}}(k_{\mbox{\tiny{rad}}})|^2
                  \sim  \frac{1}{\lambda\,\epsilon^{d-1}}\exp\left(-\frac{b}{\epsilon}\right),
\label{EnergyClassical}\end{eqnarray}
 where 
$ b\equiv2\,c_d\,\sqrt{\bar{n}^2-1}$
is an $\mathcal{O}(1)$ number. 

Now the energy of an oscillon scales as $E_{\mbox{\tiny{osc}}}\sim 1/(\lambda\,\epsilon^{d-2})$,
so the decay rate $\Gamma_d=|E_{\mbox{\tiny{osc}}}^{-1}\,dE_{\mbox{\tiny{osc}}}/dt|$ 
scales as $\Gamma_d\sim \epsilon^{-1}\exp(-b/\epsilon)$.
The constant of proportionality in $\Gamma_d$ is non-trivial to obtain. The reason is the following: at any finite order in the $\epsilon$ expansion, the corresponding terms in the equations of motion are made to vanish due to the spatial structure of the solution. However, there is always a residual piece left over which is $\mathcal{O}(\epsilon^0)$ in $k$-space, multiplied by $\phi_{\mbox{\tiny{osc}}}(k)$. 
The limit of this residual piece as we go to higher and higher order encodes the constant, but we do not pursue that here. Alternate methods for obtaining this constant can be found in the literature, such as \cite{Segur,Fodor:2009kf,Boyd1,Boyd2,Fodor:2008du}, (see also \cite{Gleiser:2008ty,Gleiser:2009ys}).
A special case is Sine-Gordon in 1-d where the remainder $J$ vanishes as $N\to\infty$, so the constant vanishes in this case.

What we have obtained in (\ref{EnergyClassical}) is the leading $\epsilon$ scaling: it is from $\bar{n}=2$ for 
asymmetric potentials and from $\bar{n}=3$ for symmetric potentials.
The form of the radiation in eq.~(\ref{EnergyClassical}) agrees with the scaling found by Segur and Kruskal \cite{Segur} in $d=1$ and generalized to other $d$. 
Our methodology of computing a spatial Fourier transform and evaluating it a wavenumber determined by kinematics is general and should apply to various other oscillon models, such as multi-field.


\section{Quantization}\label{Quantum}

The preceding discussion explains why a classical oscillon can live for an exceptionally long time.
This does, however, require the oscillon to be placed in the right initial conditions for this to occur. 
Some investigation has gone into the stability/instability of classical oscillons under arbitrary initial conditions,
e.g., \cite{Adib:2002ff,Kasuya:2002zs}. 
Here our focus is on a sharply defined question, free from the ambiguity of initial conditions: {\em If a single oscillon, as defined by the $\epsilon$ expansion, is present -- what is its lifetime?} There is a sharp answer in the classical theory -- exponential, and now we  address the question in the quantum theory.


A semi-classical (leading $\hbar$) description involves treating the
 oscillon $\phi_{\mbox{\tiny{osc}}}=\phi_{\mbox{\tiny{osc}}}(t,r)$ as a classical background and quantizing fields in this background. In this section, only the field $\phi$ itself is present to be quantized, but in Section \ref{TwoField} we will introduce a second field $\chi$ to quantize. 
  Let's write
 \begin{eqnarray}
 \phi({\bf x},t)=\phi_{\mbox{\tiny{osc}}}(x,t)+\hat{\phi}({\bf x},t),
 \end{eqnarray}
where $\hat{\phi}$ is a quantum field satisfying canonical commutation relations. At any finite order in the $\epsilon$ expansion, $\phi_{\mbox{\tiny{osc}}}$ is an exact periodic solution of the equations of motion, as discussed in Section \ref{Classical}. Perturbing around a solution allows us to write down the following equation of motion for $\hat{\phi}$ in the Heisenberg picture
\begin{eqnarray}
\ddot{\hat{\phi}}-\nabla^2\hat{\phi}+\hat{\phi} + \Phi(\phi_{\mbox{\tiny{osc}}})\hat{\phi} = 0,
\label{Heisenberg}\end{eqnarray}
where $\Phi(\phi_{\mbox{\tiny{osc}}}) \equiv V_I''\!(\phi_{\mbox{\tiny{osc}}})$.
We have neglected higher order terms in $\hat{\phi}$, since we are only interested in a leading order $\hbar$ analysis; so our results only apply in the weak coupling regime, i.e.,  $\lambda\,\hbar\ll 1$,
which is a necessary condition for the oscillon to possess a roughly classical description.
It seems reasonable, and is the common assumption in the literature on Q-balls as well as p/reheating, that for $\lambda\,\hbar\ll 1$ higher order effects will not modify the central conclusions.
Eq.~(\ref{Heisenberg}) is the theory of a free quantum scalar field with a space-time dependent mass.
The ground state of this theory is given by an infinite sum of 1-loop vacuum diagrams, coming from $N_{\Phi}=0,1,2,\ldots$ insertions of the external field $\Phi$. The $N_{\Phi}=0$ diagram corresponds to the ordinary zero point energy of a free field, while the $N_{\Phi}\geq 1$ diagrams correspond to production of $\hat{\phi}$ quanta from the background source. 
We will see, however, that although these are 1-loop effects in the semi-classical Lorentz violating theory, these processes will have an interpretation in terms of tree-level effects of the underlying Lorentz invariant theory.

 For small $\epsilon$, the oscillon is wide, with $k$-modes concentrated around $k=\mathcal{O}(\epsilon)\ll 1$,
 which suggests it is more convenient to perform the analysis in $k$-space.
So let's take the Fourier transform:
\begin{eqnarray}
\ddot{\hat{\phi}}_k+\omega_k^2\,\hat{\phi}_k+\int\frac{d^dk'}{(2\pi)^d}\Phi({\bf k}-{\bf k}')\hat{\phi}_{k'}=0,
\label{convolution}\end{eqnarray}
where $\omega_k^2\equiv k^2+1$. Here we used the convolution theorem on the final term.
Now if the background was homogeneous, then each $\hat{\phi}_k$ would decouple. 
This makes the solution rather straightforward, as is the situation during cosmological inflation, for instance.
In that case each $\hat{\phi}_k$ is proportional to a single time independent annihilation operator $\hat{a}_k$ times a mode function $v_k(t)$ that satisfies the classical equation of motion, plus hermitian conjugate.
But due to the inhomogeneity in $\phi_{\mbox{\tiny{osc}}}$, the $k$-modes are coupled, and the solution in the background of an oscillon is non-trivial. 

Nevertheless a formal solution of the Heisenberg equations of motion can be obtained. 
The key is to integrate over all annihilation operators:
\begin{eqnarray}
\hat{\phi}_k(t)=\sqrt{\hbar}\int \frac{d^dq}{(2\pi)^d}\,\hat{a}_q\,v_{qk}(t) + h.c.
\end{eqnarray}
Upon substitution, each mode function $v_{qk}(t)$ must satisfy the classical equation of motion
\begin{eqnarray}
\ddot{v}_{qk}+\omega_k^2\,v_{qk}+\int\frac{d^dk'}{(2\pi)^d}\Phi(k-k')v_{qk'}=0.
\label{Heom}\end{eqnarray}
We see that we have a matrix of time dependent mode functions $v_{qk}(t)$ to solve for. 
We choose initial conditions such that $\hat{\phi}$ is initially in its unperturbed vacuum state. This requires the following initial values of the mode functions:
\begin{eqnarray}
v_{qk}(0) &=& {1\over\sqrt{2\omega_k}}\,(2\pi)^d\,\delta^d({\bf q}-{\bf k}),\label{LOmfa}\\
\dot{v}_{qk}(0)&=& -i\,\omega_k\,v_{qk}(0).
\label{LOmf}\end{eqnarray}

The local energy density $u$ and total energy $E$ in $\hat{\phi}$ at time $t$ can be defined by the unperturbed Hamiltonian, giving
\bea
u(x,t)\amp=\amp{\hbar\over2}\int \! {d^dq\over(2\pi)^d}{d^dk\over(2\pi)^d}{d^dk'\over(2\pi)^d}\,e^{i({\bf k}'-{\bf k})\cdot{\bf x}}\nonumber\\
&&\,\,\,\,\,\,\,\,\,\,\,\,\, \times\left[\dot{v}_{qk}\dot{v}_{qk'}^*+\omega_{kk'}^2v_{qk}v_{qk'}^*\right],\,\,\,\,\,\label{EnergyD}\\
E(t)\amp=\amp{\hbar\over2}\int\! {d^dq\over(2\pi)^d}{d^dk\over (2\pi)^d}\left[ |\dot{v}_{qk}|^2+\omega_k^2|v_{qk}|^2\right]
\label{Energy}\eea
($\omega_{kk'}^2\equiv {\bf k}\cdot{\bf k}'+1$), where the second equation is easily obtained from the first by integrating over ${\bf x}$.
Initially the total energy is $E(0)=\int d^dk\frac{1}{2}\hbar\,\omega_k\,\delta^d(0)$; the usual (infinite) zero point energy of the vacuum. The energy corresponding to $\hat{\phi}$ production is contained in the time evolution of $E(t)$. As we will explain, radiation comes from specific wavenumbers that are $\mathcal{O}(1)$, and are connected to tree-level processes of the underlying microphysical theory. This makes it straightforward to extract the correct finite result for the produced radiation energy, despite the zero point energy being UV divergent.

To solve the system numerically we operate in a box of volume $V=L^d$ and discretize the system as follows
\bea
\int\frac{d^dk'}{(2\pi)^d}&\to&\frac{1}{V}\sum_{{\bf k}}
\,\,\,\,\left({\bf k}=\frac{2\pi {\bf n}}{L},\,n_i\in\mathbb{Z}\right)\!,\,\,\,\,\,\\
\delta^d({\bf q}-{\bf k})&\to& {V\over(2\pi)^d}\,\delta_{qk}.
\eea
The discretized equations of motion represent an infinite set of coupled oscillators. They have a periodic mass as driven by the background oscillon and as such are amenable to a generalized Floquet analysis of coupled oscillators. We use the Floquet theory to solve for the late time behavior, as discussed in Section \ref{Floquet}.

It is interesting to compare this to a classical stability analysis. 
Here one should explore all initial conditions that span the complete space of perturbations. To do this, we can write $\delta\phi_k\to v_{qk}$ where $q$ is merely an index that specifies the choice of initial condition. To span all initial conditions, choose $v_{qk}(0)\propto \delta({\bf q}-{\bf k})$, as in the standard Floquet theory. But this is precisely what we have done to solve the quantum problem. Hence  a classical stability analysis over all initial conditions involves the same computation as solving the quantum problem for a fixed initial condition -- the ground state. 

\section{Quantum Radiation}
\label{HigherQuantum}

Perhaps the most interesting oscillons are those that are stable against instabilities that would appear in a classical simulation; such instabilities typically arise from $k=\mathcal{O}(\epsilon)$ modes of the oscillon, and will be discussed in Section \ref{LeadingQuantum}. 
Classically, these oscillons appear to be {\em extremely} stable. In this Section we calculate and explain why the oscillon lifetime is shortened in the quantum theory due to $k=\mathcal{O}(1)$ modes, depending on the size of the effective $\hbar$.

In order to make progress, we will solve the mode functions perturbatively.
Although our methodology is general, we will demonstrate this with a model that makes the computation the easiest. Lets consider
the potential
\beq
V_I(\phi)=-{\lambda\over4!}\phi^4+{\lambda_5\over5!}\phi^5+\ldots
\eeq
The $-\lambda\,\phi^4$ term ensures that classical oscillons exist in the form given earlier (eq.~(\ref{phi1})).
The $\lambda_5\,\phi^5$ will be quite important in the quantum decay, but it does not effect the leading order term in the classical oscillon expansion. The dots represent higher order terms, such as $+\phi^6$ which ensure that the potential is well behaved at large $\phi$. It is important to note that this potential, as well as all models studied in this paper, possess oscillon solutions which are highly stable when studied classically. The only regime in which the classical oscillons are unstable depends sensitively on dimensionality and amplitude, which we will mention in Section \ref{LeadingQuantum}.

It can be verified that for this potential the oscillon has the form
\beq
\phi_{\mbox{\tiny{osc}}}(x,t)=\phi_{\epsilon}(x)\cos(\omega t)+\mathcal{O}(\epsilon^3)
\eeq
where $\phi_{\epsilon}(x)\equiv 4\,\epsilon\,\tilde{f}(x\epsilon)/\sqrt{\lambda}$ (note $\phi_\epsilon\sim\epsilon$). $\lambda_5$ only enters the expansion at $\mathcal{O}(\epsilon^4)$ (generating even harmonics). 

Lets perform an expansion in powers of $\lambda_5$, i.e.,
\beq
v_{q}=v_{q}^{(0)}+\lambda_5\,v_{q}^{(1)}+\lambda_5^2\,v_{q}^{(2)}+\ldots
\eeq
with each term $v_q^{(i)}$ implicitly an expansion in powers of $\epsilon$, with $\epsilon$ assumed small.
We have suppressed the second index on $v_{q}$, which would be $k$ in momentum space or $x$ in position space, since both representations will be informative.
In position space at zeroth order in $\lambda_5$, we have
\beq
\ddot v_{qx}^{(0)}-\nabla^2v_{qx}^{(0)}+v_{qx}^{(0)}=\left[{\lambda\over2}\phi_{\epsilon}^2(x)\cos^2(\omega t)+\mathcal{O}(\epsilon^4)\right]v_{qx}^{(0)}
\label{vq0}\eeq
The first term on the RHS is sometimes responsible for collapse instabilities, as we shall discuss in Section \ref{LeadingQuantum}, but only for wavenumbers $k=\mathcal{O}(\epsilon)$. The existence of such instabilities are highly model dependent and are not the focus of this section. Instead here we focus on $k=\mathcal{O}(1)$ which is relevant for outgoing radiation. The second term on the RHS, which is $\mathcal{O}(\epsilon^4)$, is indeed relevant to the production of radiation, but it will be superseded by radiation at $\mathcal{O}(\epsilon^3)$ that will enter when we examine $v_{q}^{(1)}$. For now, we only need to conclude from eq.~(\ref{vq0}) that $v_{q}^{(0)}$ is equal to the solution of the free-theory plus $\mathcal{O}(\epsilon^2)$ corrections.
In $k$-space this means we take the unperturbed mode functions
\beq
v_{qk}^{(0)}(t)={e^{-i\omega_k t}\over\sqrt{2\omega_k}}(2\pi)^d\delta^d({\bf q}-{\bf k}) +\mathcal{O}(\epsilon^2),
\eeq
which match the initial conditions mentioned earlier in eqs.~(\ref{LOmfa},\,\ref{LOmf}). Note that an integral over $q$ recovers the standard mode functions that occur in a free theory.

At next order in $\lambda_5$ we have 
\beq
\ddot v_{qx}^{(1)}-\nabla^2v_{qx}^{(1)}+v_{qx}^{(1)}=-{1\over3!}\phi_{\epsilon}^3(x)\cos^3(\omega t)v_{qx}^{(0)}
\label{vq1}\eeq
Fourier transforming to $k$-space and inserting $v_{qk}^{(0)}$ gives
\bea
\ddot v_{qk}^{(1)}+\omega_k^2v_{qk}^{(1)}
\amp = \amp -{\psi_{\epsilon}({\bf q}-{\bf k})\over3!2^3\sqrt{2 \omega_k}}\left[e^{i(\tilde\omega-\omega_q) t}+e^{-i(\tilde\omega+\omega_q) t}\right]\,\,\,\,\,\,\,\,\,\,\,
\eea
where 
$\psi_\epsilon(k)$  
is the Fourier transform of $\phi_\epsilon^3(x)$ and $\tilde\omega\equiv3\,\omega$.
Here we have ignored terms on the RHS with frequency $\omega\pm\omega_q$ which cannot generate a resonance.
The first term included does generate a resonance at $\omega_q\approx\omega_k\approx\tilde\omega/2$.
This is the equation of a forced oscillator whose solution can be readily obtained. By imposing the initial conditions
$v_{qk}^{(1)}(0)=\dot v_{qk}^{(1)}(0)=0$ we find
\bea
v_{qk}^{(1)}(t)=-{\psi_\epsilon({\bf q}-{\bf k})\over3!2^3\sqrt{2\omega_k}}
S(\tilde\omega,\omega_q,\omega_k)
\eea
where
\bea
 S(\tilde\omega,\omega_q,\omega_k\!\!\!\amp)\amp\!\equiv
{e^{-i(\tilde\omega+\omega_q)t}\over-(\omega_q+\tilde\omega)^2+\omega_k^2}
+{e^{i(\tilde\omega-\omega_q)t}\over-(\omega_q-\tilde\omega)^2+\omega_k^2}\nonumber\\
\amp-\amp {\left(1-{\omega_q\over\omega_k}\right)e^{-i\omega_k t}\over-\tilde\omega^2+(\omega_k-\omega_q)^2}
-{\left(1+{\omega_q\over\omega_k}\right)e^{i\omega_k t}\over-\tilde\omega^2+(\omega_k+\omega_q)^2}\,\,\,\,\,\,\,\,\,\,\,\,
\label{Sfunc}\eea

Lets now insert our solution for $v_{qk}$ into the expression for the energy density $u(x,t)$ (eq.~(\ref{EnergyD})).
This includes terms scaling as $v_{qk}^{(0)}v_{qk'}^{(0)}$ which is the usual zero point energy.
Another term scales as $\lambda_5\,v_{qk}^{(0)}v_{qk'}^{(1)}$ which is non-resonant. Then there are two important terms:
$\lambda_5^2\,v_{qk}^{(1)}v_{qk'}^{(1)}$ and $\lambda_5^2\,v_{qk}^{(0)}v_{qk'}^{(2)}$. It can be shown that the second term here provides the same contribution as the first term. (In the case of a homogeneous pump field the mode functions are often written in terms of Bogoliubov coefficients, which make this fact manifest. 
A similar argument goes through in our more complicated inhomogeneous case.)
 Given this, we will not write out the explicit solution for 
$v_{qk}^{(2)}$ here. Finally, we shall use the fact that only terms near resonance contribute significantly to the integral,
and further we shall make simplifications using $\omega_q\approx\omega_k\approx\tilde\omega/2$ whenever we can; however, this must not be done in singular denominators or in the arguments of oscillating functions.

Altogether at leading order in $\lambda_5$ and $\epsilon$ we find the following expression for the change in the energy density from the zero point:
\begin{widetext}
\bea
\delta u(x,t)={\lambda_5^2\hbar\over (3!)^2 2^8}\int \amp \amp\!\!\!\!\!\!\!\!
{d^dq\over(2\pi)^d}{d^dk\over(2\pi)^d}{d^dk'\over(2\pi)^d}\,\cos(({\bf k}'-{\bf k})\cdot{\bf x})\,\psi_\epsilon({\bf q}-{\bf k})\psi_\epsilon^*({\bf q}-{\bf k}')\nonumber\\
\amp\times\amp\left[{1+\cos(t(\omega_{k'}-\omega_k))-\cos(t(\omega_k+\omega_q-\tilde\omega))
-\cos(t(\omega_{k'}+\omega_q-\tilde\omega))
\over\omega_q(\omega_k+\omega_q-\tilde\omega)(\omega_{k'}+\omega_q-\tilde\omega)}\right].
\label{densityresult}\eea
\end{widetext}

\newpage

\newpage

\begin{figure}[b]
\center{\includegraphics[width=\columnwidth]{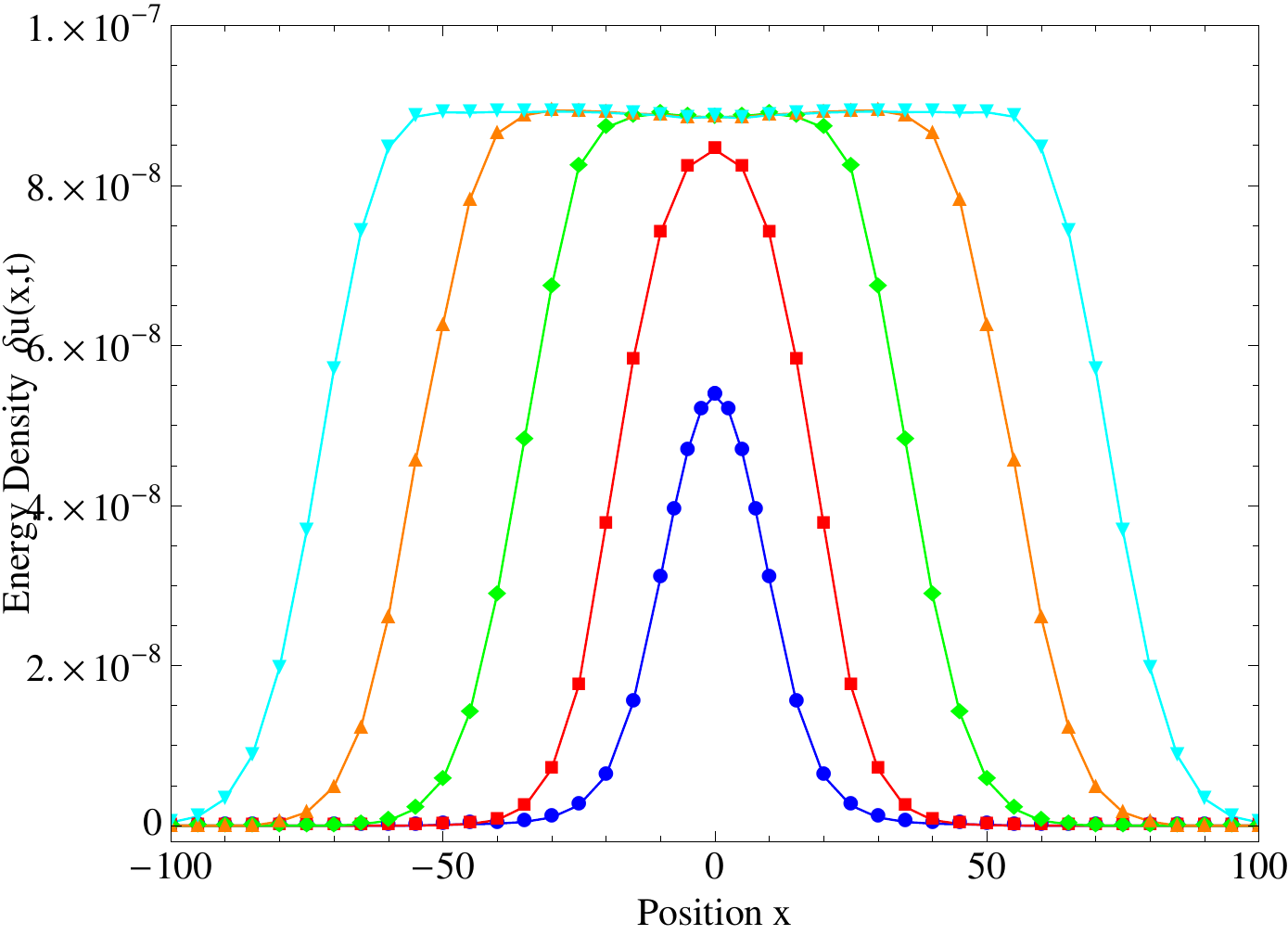}}
\caption{
Energy density $\delta u(x,t)$ (in units of $\hbar$). Each curve is at a different time interval:
blue is $t=10$, red is $t=25$, green is $t=50$, orange is $t=75$, and cyan is $t=100$.
Here $d=1$, $\epsilon=0.05$, and $\lambda_5^2/\lambda^3=1$.}
\label{1fieldEnergyOut}
\end{figure}

For the 1-dimensional case we have evaluated this numerically; the results are plotted in Fig.~(\ref{1fieldEnergyOut}).
As the figure shows the energy density initially grows in the core of the oscillon before reaching a maximum (around $\delta u\sim 10^{-7}$ in the figure for the parameters chosen). This occurs at each point in space and so the net effect is of continual growth in $\delta u$ by spreading out. This has a simple interpretation: $\phi$-particles are being produced from the core of the oscillon and moving outwards. We will show using scattering theory that this is connected to the annihilation process $3\,\phi\to2\,\phi$.

Let us turn now to the total energy output (after subtracting the zero point) $\delta E(t)$. This comes from integrating 
eq.~(\ref{densityresult}) over ${\bf x}$, giving
\bea
\delta E(t)={\lambda_5^2\hbar\over (3!)^22^7}\int \amp \amp\!\!\!\!\!\!\!\!
{d^dq\over(2\pi)^d}{d^dk\over(2\pi)^d}|\psi_\epsilon({\bf q}-{\bf k})|^2\nonumber\\
\amp\times\amp\left[{1-\cos(t(\omega_k+\omega_q-\tilde\omega))
\over\omega_q(\omega_k+\omega_q-\tilde\omega)^2}\right]\!.\,\,\,\,\,\,\,\,
\label{Energynotsimp}\eea
This may be further simplified by recognizing that $|\psi_\epsilon({\bf q}-{\bf k})|^2$ is non-negative and sharply spiked around ${\bf q}-{\bf k}\approx 0$ for small $\epsilon$ (as we discussed in Section \ref{ExpSmall}) and so it acts like a $\delta$-function, i.e.,
\beq
|\psi_\epsilon({\bf q}-{\bf k})|^2\approx \left[\int{d^dk\over(2\pi)^d}|\psi_\epsilon({\bf k})|^2\right]
(2\pi)^d\delta^d({\bf q}-{\bf k}),
\label{deltarep}\eeq
where we have intoduced the integral prefactor
to ensure that the integration of $|\psi_\epsilon({\bf q}-{\bf k})|^2$ over all ${\bf k}$ gives the correct value when we replace it by the $\delta$-function. This quantity has a nice interpretation when re-written as an integral over position space. 
Since $\psi_\epsilon(k)$ is the Fourier transform of $\phi_\epsilon^3(x)$, we can write
\bea
\int{d^dk\over(2\pi)^d}|\psi_\epsilon(k)|^2\approx 2^3\!\int d^dx\,u_\epsilon^3(x),
\eea
where $u_\epsilon(x)\approx \phi_\epsilon^2(x)/2$ is the oscillon's local energy density.
Inserting this into (\ref{Energynotsimp}) allows us to write the energy output in terms of an integral over the occupancy number $n_k(t)$ as follows
\beq
\delta E(t)=\int \! {d^dk\over(2\pi)^d}\hbar\,\omega_k\,n_k(t)
\label{Energysimp}\eeq
with
\bea
n_k(t)={\lambda_5^2\int d^dx\,u_\epsilon^3(x)\over (3!)^22^4}{1-\cos(t(2\omega_k-\tilde\omega))
\over\omega_k^2(2\omega_k-\tilde\omega)^2}.\,\,\,
\eea
Note that the occupancy number is resonant at $\omega_k\approx\tilde\omega/2$.
The late time behavior can be studied by noting that as $t\to\infty$
\beq
{1-\cos(t(2\omega_k-\tilde\omega))  \over(2\omega_k-\tilde\omega)^2}\to{\pi\,t\over2}
\delta(\omega_k-\tilde\omega/2).
\label{deltapres}\eeq
This shows an important connection; our results are a form of {\em Fermi's golden rule}, where the 
delta-function prescription is always used and is known to provide accurate transition rates.

Note that no other form of ``renormalization" is required here, since the process we are computing is connected to tree-level processes of the underlying Lorentz invariant quantum field theory (albeit 1-loop in the semi-classical Lorentz violating theory) and not loop-level processes which are higher order in the couplings and $\hbar$. We will elaborate on this connection shortly.

To compute the energy output in the Fermi golden rule approximation, we begin by
writing $d^dk=d\Omega\,dk\,k^{d-1}$. The angular integration in (\ref{Energysimp})
is trivial $\int d\Omega=2\,\pi^{d/2}\!/\Gamma(d/2)$. Now write $dk=d\omega_k \omega_k/k$. The integral over $d\omega_k$ is trivial due to the delta-function (\ref{deltapres}), which enforces 
$k$ to take on its resonant value corresponding to radiation $k_{\mbox{\tiny{rad}}}$, which satisfies
$\omega_k=\sqrt{k_{\mbox{\tiny{rad}}}^2+m^2}=3\,\omega/2$.
This gives the following expression for the energy output in the Fermi golden rule approximation
\bea
\delta E(t)={\lambda_5^2\hbar\over (3!)^22^4}{\pi^{d/2+1}k_{\mbox{\tiny{rad}}}^{d-2}\over\Gamma({d\over2})(2\pi)^d}
\int d^dx\,u_\epsilon^3(x)\,t.
\label{EnergyLinear}\eea

Hence the energy output increases linearly with time. So the oscillon must lose energy at this rate.
The decay rate is $\Gamma_d=|E_{\mbox{\tiny{osc}}}^{-1}\,dE_{\mbox{\tiny{osc}}}/dt|$ and using $E_{\mbox{\tiny{osc}}}=\int d^dx\,u_\epsilon(x)$ we obtain
\beq
\Gamma_d(3\,\phi\to2\,\phi)= {\lambda_5^2\hbar\over (3!)^22^4}{\pi^{d/2+1}k_{\mbox{\tiny{rad}}}^{d-2}\over\Gamma({d\over2})(2\pi)^d}
{\int d^dx\,u_\epsilon^3(x)\over \int d^dx\,u_\epsilon(x)}
\label{DecayFinal}\eeq
as our final result for the decay rate. We have labelled this ``$3\,\phi\to2\,\phi"$ annihilation for a reason we now explain.

Its useful to connect this result to ordinary perturbation theory for a gas of incoherent particles. The differential transition rate for a single annihilation of $N_i\,\phi\to 2\,\phi$ in a box of volume $V_{\mbox{\tiny{box}}}$, evaluated on threshold for non-relativistic initial particles of mass $m$, is given by (e.g., see \cite{WeinbergVol1})
\begin{eqnarray}
d\Gamma_1\!=\!\frac{V_{\mbox{\tiny{box}}}^{1-N_i}}{N_i!(2m)^{N_i}} |\mathcal{M}|^2
(2\pi)^{D}\delta^{D}\!(p_i-p_f) \! \prod_f \! \frac{d^d p_f}{(2\pi)^d}\frac{1}{2 E_f}\,\,\,
\label{Scattheory}\end{eqnarray}
($D=d+1$).
To connect to the previous calculation we choose $N_i=3$ since the relevant interaction is provided by $3\,\phi\to2\,\phi$ annihilation at tree-level due to the interaction term $\Delta V_I={1\over5!}\lambda_5\,\phi^5$.
This gives $|\mathcal{M}|^2=\lambda_5^2$. Performing the integration over phase space (including division by 2 since the final 2 particles are indistinguishable) gives the result
\beq
\Gamma_1(3\,\phi\to2\,\phi)= {\lambda_5^2\hbar\over 3! 2^4}{\pi^{d/2+1}k_{\mbox{\tiny{rad}}}^{d-2}\over\Gamma({d\over2})(2\pi)^d}{V_{\mbox{\tiny{box}}}^{-2}\over2E_f}
\label{Gam1}\eeq
The inverse of this is the time taken for a given triplet of $\phi$'s to annihilate. For a box of $N_{\phi}$ particles the total rate for an annihilation to occur is $\Gamma_1$ multiplied by the number of indistinguishable ways we can choose a triplet, i.e.,
\bea
\Gamma_{\mbox{\tiny{tot}}}(3\,\phi\to2\,\phi)\amp=\amp
\left(\!\begin{array}{c}N_\phi \\ 3\end{array}\!\right) \Gamma_1(3\,\phi\to2\,\phi)\\
\amp\approx\amp{N_\phi^3\over 3!}\,\Gamma_1(3\,\phi\to2\,\phi)\,\,\,\,
\eea
for $N_\phi\gg 1$ (the semi-classical regime).
Since each process produces a pair of particles of total energy $2E_f$, the total energy output is
\beq
\delta E(t)=2E_f\Gamma_{\tiny{\mbox{tot}}}(3\,\phi\to2\,\phi)\, t. 
\eeq
Together with (\ref{Gam1}) this recovers the result in eq.~(\ref{EnergyLinear}) precisely,
as long as we make the replacement
\beq
V_{\mbox{\tiny{box}}}^{-2} N_{\mbox{\tiny{box}}}^3\to \int d^dx\, u_\epsilon^3(x).
\label{Connection}\eeq

Hence to leading order in $\lambda_5$ and $\epsilon$ we find that decay rates computed using tree-level scattering theory recover
the result computed from solving the mode functions for a coherent background oscillon.  The important connection
is provided by eq.~(\ref{Connection}) to account for the spatial structure of the oscillon.

\begin{figure}[t]
\includegraphics[width=3cm,height=2.5cm, angle=90]{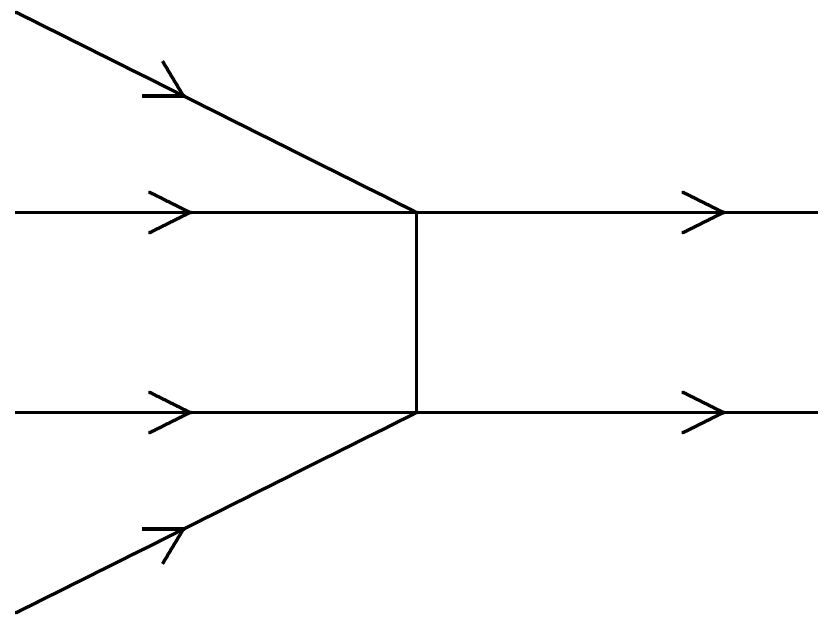}\,\,\,\,\,\, 
 \includegraphics[width=3cm,height=2.5cm,angle=90]{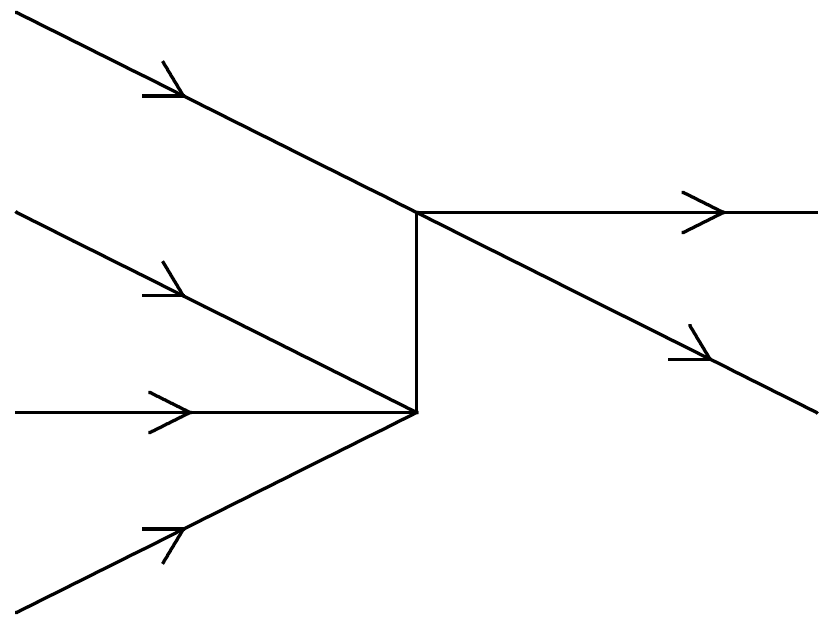}\,\,\,\,\,\, 
  \includegraphics[width=3cm,height=2.5cm,angle=90]{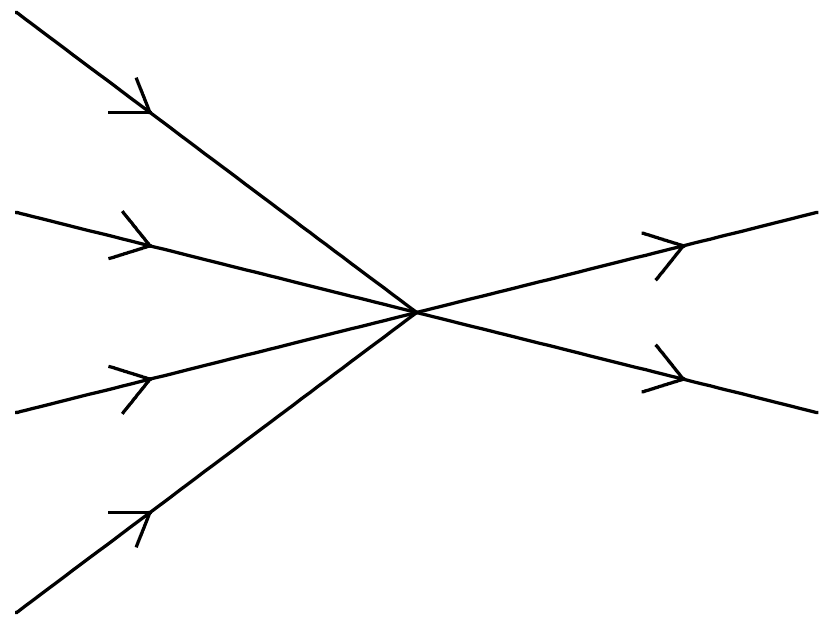} 
\caption{Feynman diagrams for the process $4\,\phi\to 2\,\phi$. 
Evaluated on threshold, the first diagram is $i\lambda_4^2/4$ ($\times\,6$ crossing symmetries), the second diagram 
$-i\lambda_4^2/8$ ($\times\,4$ crossing symmetries), and the third diagram is $-i\lambda_6$.}
\label{Feynman}
\end{figure}

Given this connection, the generalization of our result to arbitrary interactions is relatively straightforward.
Consider the more general interaction potential
\beq
V_I(\phi)={\lambda_3\over3!}\phi^3+{\lambda_4\over 4!}\phi^4+{\lambda_5\over 5!}\phi^5+{\lambda_6\over 6!}\phi^6+\ldots
\eeq
As mentioned earlier, a requirement for the existence of small amplitude oscillons is
$\frac{5}{3}\lambda_3^2-\lambda_4>0$; this in fact is the {\em only} requirement on the couplings \cite{Fodor:2008du}.
If either $\lambda_3$ or $\lambda_5$ are non-zero, then the leading order radiation arises from $3\,\phi\to2\,\phi$ annihilation,
as we computed previously for the $\lambda_3=0$ case. This again gives the result of eq.~(\ref{DecayFinal}), with the generalization
of the $\lambda_5^2$ prefactor being replaced by the square of the matrix element for all such tree-level scattering diagrams, i.e.,
\beq
\lambda_5^2\to|\mathcal{M}(3\,\phi\to 2\,\phi)|^2
\label{GammaGeneral}\eeq
This includes the important case in $d=3$ with $\lambda_3> 0$, $\lambda_4> 0$ and $\lambda_5=\lambda_6=\ldots=0$, which is a renormalizable field theory. For brevity, we have not drawn the many relevant diagrams here, but in Fig.~\ref{Feynman} we draw the diagrams that are relevant to the next example.

Now consider a purely symmetric potential with $\lambda_3=\lambda_5=0$ and $\lambda_4<0$ and $\lambda_6>0$.
 In this case, the leading order annihilation process is $4\,\phi\to 2\,\phi$, so we must choose $N_i=4$ in eq.~(\ref{Scattheory}). By summing the diagrams of Fig.~\ref{Feynman} we find $|\mathcal{M}(4\,\phi\to2\,\phi)|^2=(\lambda_4^2-\lambda_6)^2$. Altogether we obtain the following result
\beq
\Gamma_d(4\,\phi\to2\,\phi)= {(\lambda_4^2-\lambda_6)^2\hbar\over (4!)^22^5}{\pi^{d/2+1}k_{\mbox{\tiny{rad}}}^{d-2}\over\Gamma({d\over2})(2\pi)^d}
{\int d^dx\,u_\epsilon^4(x)\over \int d^dx\,u_\epsilon(x)}\,\,\,\,\,\,\,
\label{DecayFinal2}
\eeq
In this case the kinematical requirement on $k_{\mbox{\tiny{rad}}}$ for resonance is 
$\omega_k=\sqrt{k_{\mbox{\tiny{rad}}}^2+m^2}=2\,\omega$.
Note this rate vanishes if and only if $\lambda_6=\lambda_4^2$, 
which is true for the Sine-Gordon potential $V_{\mbox{\tiny{SG}}}(\phi)=(1-\cos(\sqrt{\lambda}\,\phi))/\lambda$. In fact it can be proven that in (1+1)-dimensions every $\phi$ number changing process vanishes on threshold for the Sine-Gordon potential.

Lets discuss the scaling of our results (\ref{DecayFinal}) and 
({\ref{DecayFinal2}). Recall that $\phi_{\mbox{\tiny{osc}}}\sim\epsilon/\sqrt{\lambda}$, so $u_\epsilon\sim\epsilon^2/\lambda$. 
Thus (depending on the relative size of couplings) the decay rates scale as
\bea
&&\Gamma_d(3\,\phi\to2\,\phi)\sim\epsilon^4\lambda_5^2\hbar/\lambda^2,\,\,\,\,\mbox{or}\,\,\,\,\,\,\epsilon^4\lambda_3^6\hbar/\lambda^2\,\,\,\,\,\,\,\,\\
&&\Gamma_d(4\,\phi\to2\,\phi)\sim\epsilon^6\lambda_4\hbar,\,\,\,\,\,\,\,\,\,\,\,\,\,\mbox{or}\,\,\,\,\,\,\epsilon^6\lambda_6^2\hbar/\lambda^3.
\eea 
(We also find that in the large $+\lambda_6\,\phi^6$ model of Ref.~\cite{Mustafa} the scaling is $\Gamma_d\sim \lambda_4^3\hbar/\lambda_6$.)
Hence the quantum mechanical decay rate is a power law in the parameters $\epsilon,\, \lambda_i$. 
Such quantum decay rates will only be smaller than the corresponding classical decay rate (\ref{EnergyClassical})
if the ``effective $\hbar$" (such as $\lambda\,\hbar$) is {\em extremely} small, as we are comparing it to an exponentially small quantity in $\exp(-b/\epsilon)$.
Unlike collapse instabilities that we will mention in Section \ref{LeadingQuantum}, it is almost impossible to avoid this radiation by changing dimensionality, parameter space, or field theory (the only known exception is the Sine-Gordon model in $d=1$).

One may wonder why the classical analysis failed to describe this decay rate accurately given that the oscillon
can be full of many quanta, say $N_\phi\approx M_{\mbox{\tiny{osc}}}/m_\phi$, with an almost continuous spectrum. Let us explain the resolution.
For $N_\phi\gg1$ the quantum corrections to 
the oscillon's bulk properties are small.
For example, the quantum correction to the oscillon's width, amplitude, total mass, etc, should be small, since the
classical values are large. In $k$-space we can say that these properties are governed by $k=\mathcal{O}(\epsilon)$ modes,
which carry a large amplitude.
However, the radiation is quite different; it is governed by $k=\mathcal{O}(1)$ modes, which are exponentially small in the classical oscillon, but are not exponentially small in the quantum oscillon due to zero point fluctuations.
Another way to phrase this is to consider the commutation relation
\beq
\hat{\phi}({\bf k})\hat{\pi}({\bf k}')= \hat{\pi}({\bf k}')\hat{\phi}({\bf k})+ i\hbar\,(2\pi)^d\delta^d({\bf k}-{\bf k}').
\eeq
For $k=\mathcal{O}(\epsilon)$ the classical value of the LHS and its counterpart on the RHS are both large, so the $\hbar$ correction is negligible. But for $k=\mathcal{O}(1)$ the classical values are small, so the quantum corrections are important.

An interesting issue is the behavior of the growth at late times. In the case of a homogeneous background pump field (for example, as is relevant during p/reheating at the end of inflation, e.g., see \cite{Kofman:1997yn,Yoshimura:1995gc,Yoshimura:1996fk}) it is known that the linear growth is really the initial phase of exponential growth.
So is the same true for spatially localized oscillons? Here we claim that  for sufficiently small amplitude oscillons, the answer is {\em no}, the linear growth rate is correct even at late times, although this can change for sufficiently large amplitude oscillons where our perturbative analysis breaks down. We have confirmed the linear growth for small amplitude oscillons in 2 ways: (i) by expanding $v_{qk}$ to higher order in the coupling and (ii) by solving the full mode function equations (\ref{Heom}) numerically.  
The reason for this result is subtle and will be discussed in detail in Section \ref{ExpLinear}; we will demonstrate that whether the growth is in the linear regime or exponential regime depends critically on the oscillon's amplitude, width, and couplings.

\section{Coupling to Other Fields}\label{TwoField}

Most fields in nature interact considerably with others.
It is important to know what is the fate of a $\phi$-oscillon that is coupled to other fields.
Let's couple $\phi$ to another scalar $\chi$ and consider the following Lagrangian
\bea
\mathcal{L}\amp=\amp\frac{1}{2}(\partial\phi)^2-\frac{1}{2}m^2\,\phi^2-V_I(\phi)
+\frac{1}{2}(\partial\chi)^2-\frac{1}{2}m_\chi^2\,\chi^2\nonumber\\
\amp\amp-{1\over 2}g_1\,m\,\phi\,\chi^2-{1\over 2}\,g_2\,\phi^2\chi^2.
\eea
The last two terms represent interactions between the 2 fields, with $g_i$ coupling parameters. 
The interaction term $g_1\,\phi\,\chi^2$ allows the following tree-level decay process to occur in vacuo:
$\phi\to\chi+\chi$,
if the following mass condition is met:
$m>2\,m_\chi$.
Assuming this condition is met, we are led to ask: Will the $\phi$-oscillon decay into $\chi$? 
If so, will the growth in $\chi$ be linear or exponential?
Otherwise, if $m<2\,m_\chi$, or if $g_1=0,\,g_2\neq 0$, we can focus on annihilations: $\phi+\phi\to\chi+\chi$, etc.


One could approach the issue of multiple fields by evolving the full $\phi$, $\chi$ system under the classical equations of motion. 
In fact by scanning the mass ratio of $\phi$ and $\chi$ one can find interesting oscillons involving an interplay of both fields -- one finds that the 2:1 mass ratio is of particular importance (as was the case for the SU(2) oscillon \cite{Farhi:2005rz}).
Although this is an interesting topic, here we would like to focus on the effects on the $\phi$-oscillon due to the introduction of $\chi$, initially in its vacuum state. At the classical level $\chi$ will remain zero forever, so this is trivial. We will return to the issue of the classical evolution for non-trivial initial conditions for $\chi$ in the next Section. For now we focus on placing $\chi$ in its quantum vacuum state with a classical background $\phi_{\mbox{\tiny{osc}}}$.

We use the same formalism as we developed in Section \ref{Quantum}.
We make the replacements $\hat{\phi}\to\hat{\chi}$, $m=1\to m_\chi$ in eq.~(\ref{convolution}), giving the following Heisenberg equation of motion for $\hat{\chi}$ in $k$-space
\begin{eqnarray}
\ddot{\hat{\chi}}_k+\omega_k^2\,\hat{\chi}_k+\int\frac{d^dk'}{(2\pi)^d}\Phi({\bf k}-{\bf k}')\hat{\chi}_{k'}=0,
\label{convolution2}\end{eqnarray}
where $\omega_k^2\equiv k^2+m_\chi^2$ and $\Phi\equiv g_1\,\phi_{\mbox{\tiny{osc}}}({\bf x},t)+g_2\,\phi_{\mbox{\tiny{osc}}}^2({\bf x},t)$.
We write $\hat{\chi}$ in terms of its mode functions $v_{qk}$ as before
\begin{eqnarray}
\hat{\chi}_k(t)=\sqrt{\hbar}\int \frac{d^dq}{(2\pi)^d}\,\hat{a}_q\,v_{qk}(t) + h.c.
\end{eqnarray}
For brevity, lets focus on leading order in $\epsilon$ behavior, coming from $\Phi\approx g_1\, \phi_\epsilon(x)\cos(\omega t)$ (although later we will also mention the important case of $g_1=0$ with $\Phi=g_2\,\phi_\epsilon^2(x)\cos^2(\omega t)$).
This gives the following mode function equations
\beq 
\ddot v_{qk}+\omega_k^2v_{qk}+g_1\cos(\omega t)\int{d^dk'\over(2\pi)^d}\,\phi_\epsilon({\bf k}-{\bf k}')\,v_{qk'} = 0.
\eeq

For small coupling $g_1$ we expect the solutions of the mode equations to be small deformation of planes waves.
To capture this, lets expand the mode functions in powers of $g_1$ (analogously to the earlier expansion in Section \ref{HigherQuantum})
\beq
v_{qk}=v_{qk}^{(0)}+g_1\,v_{qk}^{(1)}+g_1^2v_{qk}^{(2)}+\ldots
\eeq
At leading order $\mathcal{O}(g_1^0)$, we have
$\ddot v_{qk}^{(0)}+\omega_k^2 v_{qk}^{(0)}=0$,
whose desired solution is the unperturbed mode functions
\beq
v_{qk}^{(0)}(t)={e^{-i\omega_k t}\over\sqrt{2\omega_k}}(2\pi)^d\delta^d({\bf q}-{\bf k}),
\eeq
as earlier. At next order we have the following forced oscillator equation
\bea
\ddot v_{qk}^{(1)}+\omega_k^2v_{qk}^{(1)} = -{\phi_{\epsilon}({\bf q}-{\bf k})\over2\sqrt{2 \omega_k}}\left[e^{-i(\omega+\omega_q) t}+e^{i(\omega-\omega_q) t}\right]\!,
\,\,\,\,\,\,
\eea
The solution with boundary conditions $v_{qk}^{(1)}(0)=\dot v_{qk}^{(1)}(0)=0$ is
\bea
v_{qk}^{(1)}(t)=-{\phi_\epsilon({\bf q}-{\bf k})\over\sqrt{2\omega_k}}
S(\omega,\omega_q,\omega_k)
\eea
where $S$ was defined in eq.~(\ref{Sfunc}).
Hence we obtain the same expressions for the energy density $\delta u$ and $\delta E$ as in eqs.~(\ref{densityresult},\,\ref{Energynotsimp}) upon making the replacements
\beq
{\lambda_5\over 3!2^3}\psi_\epsilon({\bf q}-{\bf k})\to{g_1\over2}\phi_\epsilon({\bf q}-{\bf k}),\,\,\,\,\,\,\,\,\tilde\omega\to\omega
\eeq
Evaluating this we find results that are qualitatively similar to before: the $\chi$ field is produced in
the core of the oscillon and spreads out, and the energy grows linearly in time.

Following through a similar calculation to before in the small $\epsilon$ limit, by identifying $|\phi_\epsilon({\bf q}-{\bf k})|^2$ as proportional to a $\delta$-function, i.e.,
\beq
|\phi_\epsilon({\bf q}-{\bf k})|^2\approx 2\left[\int d^dx\,u_\epsilon(x)\right]\!(2\pi)^d\delta^d({\bf q}-{\bf k}),
\eeq
allows us to evaluate the decay rate at leading order in $g_1$, which is connected to the decay process $\phi\to\chi+\chi$. We also carry though the calculation at leading order in $g_2$ (including next order in $g_1$), which is connected to the annihilation process $\phi+\phi\to\chi+\chi$. We find the results
\bea
\Gamma_d(1\,\phi\to2\,\chi) \amp = \amp {g_1^2\hbar\over2^2} {\pi^{d/2+1}k_{\mbox{\tiny{rad}}}^{d-2}\over\Gamma({d\over2})(2\pi)^d}\label{Weak12},\\
\Gamma_d(2\,\phi\to2\,\chi) \amp = \amp {(g_1^2-g_2)^2\hbar\over 2^3} {\pi^{d/2+1}k_{\mbox{\tiny{rad}}}^{d-2}\over\Gamma({d\over2})(2\pi)^d}
{\int d^dx\,u_\epsilon^2(x)\over\int d^dx\,u_\epsilon(x)}\,\,\,\,\,\,\,\,\,\,\label{Weak22}
\eea
In (\ref{Weak12}) the kinematical requirement on $k_{\mbox{\tiny{rad}}}$ is $\omega_k=k_{\mbox{\tiny{rad}}}^2+m_\chi^2=(\omega/2)^2$
and in (\ref{Weak22}) the requirement is $\omega_k^2=k_{\mbox{\tiny{rad}}}^2+m_\chi^2=\omega^2$.
So these decays only occur for sufficiently light $\chi$.
Here $\Gamma_d(1\,\phi\to2\,\chi)$ coincides with the perturbative decay rate of $\phi$ and
$\Gamma_d(2\,\phi\to2\,\chi)$ is a generalization of the annihilation rate $\Gamma=n\,\langle\sigma v\rangle$ applied to a gas of non-relativistic particles with variable density.
As an application, we expect this result to be relevant to the bosonic SU(2) oscillon \cite{Farhi:2005rz}.
Since it exists at the 2:1 mass ratio ($m_{\mbox{\tiny{H}}}=2m_{\mbox{\tiny{W}}}$), it prevents Higgs decaying into W-bosons, but it should allow the quantum mechanical annihilation of Higgs into relativistic W-bosons.

\section{Exponential vs Linear Growth}\label{ExpLinear}

So far we have worked to leading order in the couplings and $\epsilon$, this has resulted in
a constant decay rate of the oscillon into quanta of $\phi$ in the single field case or quanta of $\chi$ in the two field case.
These quanta have an energy that grows linearly in time, see eq.~(\ref{EnergyLinear}). However, one may question whether
this result applies at late times. In the case of a homogeneous background pump field, it is always the case that the growth is exponential at late times if the daughter field is bosonic. This is due to a build up in the occupancy number in certain $k$-modes, leading to rapid growth for fields satisfying Bose-Einstein statistics. 

\subsection{Floquet Analysis}\label{Floquet}

Such exponential growth is obtained by fully solving the mode functions non-perturbatively \cite{Kofman:1997yn}.
Since the background is periodic, the mode functions satisfy a form of Hill's equation, albeit with infinitely many coupled oscillators due to the spatial structure of the oscillon. There is a large literature on resonance from homogeneous backgrounds, particularly relevant to inflation, but rarely are inhomogeneous backgrounds studied as we do here.

The late time behavior is controlled by Floquet exponents $\mu$. To make this precise, consider the discrete (matrix) version of the mode function equations
\bea
\dot w_{qk} \amp = \amp z_{qk}\\
\dot{z}_{qk} \amp =\amp \sum_{k'}Q_{kk'}(t)w_{qk'}
\eea
where 
\beq
Q_{kk'}(t)\equiv -\omega_k^2\delta_{kk'}-{1\over V}\Phi({\bf k}-{\bf k}',t)
\eeq
is a matrix in $k$-space, with period $T=2\pi/\omega$. Here we have labelled the mode functions
$w$ instead of $v$, since we will impose slightly different initial conditions on $w$.
In particular consider the following pair of (matrix) initial conditions
\bea
(i) \,\,\,\,\,\,\,\, && w_{qk}(0)=\delta_{qk},\,\,\,\,z_{qk}(0)=0,\\
(ii) \,\,\,\,\,\,\,\, && w_{qk}(0)=0,\,\,\,\,\,\,\,\,\,z_{qk}(0)=\delta_{qk}.\,\,\,\,\,\,\,\,\,\,\,\,
\eea
Let $N$ be the number of $q,k$ values in our discretization.
Lets organize this information into a $2N\times 2N$ matrix $M(t)$,
whose upper left quadrant is $w_{qk}$ with IC (i), upper right quadrant is $z_{qk}$ with IC (i),
lower left quadrant is $w_{qk}$ with IC (ii), and lower right quadrant is $z_{qk}$ with IC (ii).
So initially we have $M(0)=\mathbf{1}_{2N,2N}$.
Numerically, we evolve this through one period, giving the matrix  $M(T)$.
After $n$ oscillations, we have $M(nT)=M(T)^n$.
Hence the matrix $M(T)$ controls the behavior of the system.  To obtain the result for the initial conditions of (\ref{LOmfa},\,\ref{LOmf}) we multiply $M(nT)$ onto the following diagonal matrix
\beq
\left(\begin{array}{cc}
{1\over{\sqrt{2\omega_k}}}V\delta_{kq} & 0\\
0 & {-i\omega_{\tilde k}\over{\sqrt{2\omega_{\tilde{k}}}}}V\delta_{\tilde{k}\tilde{q}}
\end{array}\right).
\eeq

The existence of exponential growth is governed by the eigenvalues of $M(T)$,
with some corresponding eigenvector $\{w_k,\,z_k\}$. Following the standard Floquet theory,
we write the eigenvalues as $\exp(\mu T)$, where $\mu$ are the Floquet exponents,
which are in general complex.
Note that although we have phrased this in the context of solving the quantum problem,
it is also a classical stability analysis, as we mentioned at the end of Section \ref{Quantum}.

\subsection{Results}

We have carried out the numerical analysis for different models, but would like to report on the
results for the ${1\over2}g_1\phi\,\chi^2$ theory with $\Phi\approx g_1\,\phi_\epsilon(x)\cos(\omega t)$ in $d=1$.
The numerical results for the maximum value of the real part of 
$\mu_{\mbox{\tiny{max}}}$ as a function of $g_1$ is given in Fig.~\ref{quantumgrowth} (top panel, red curve).
\begin{figure}[t]
\center{ \includegraphics[width=\columnwidth]{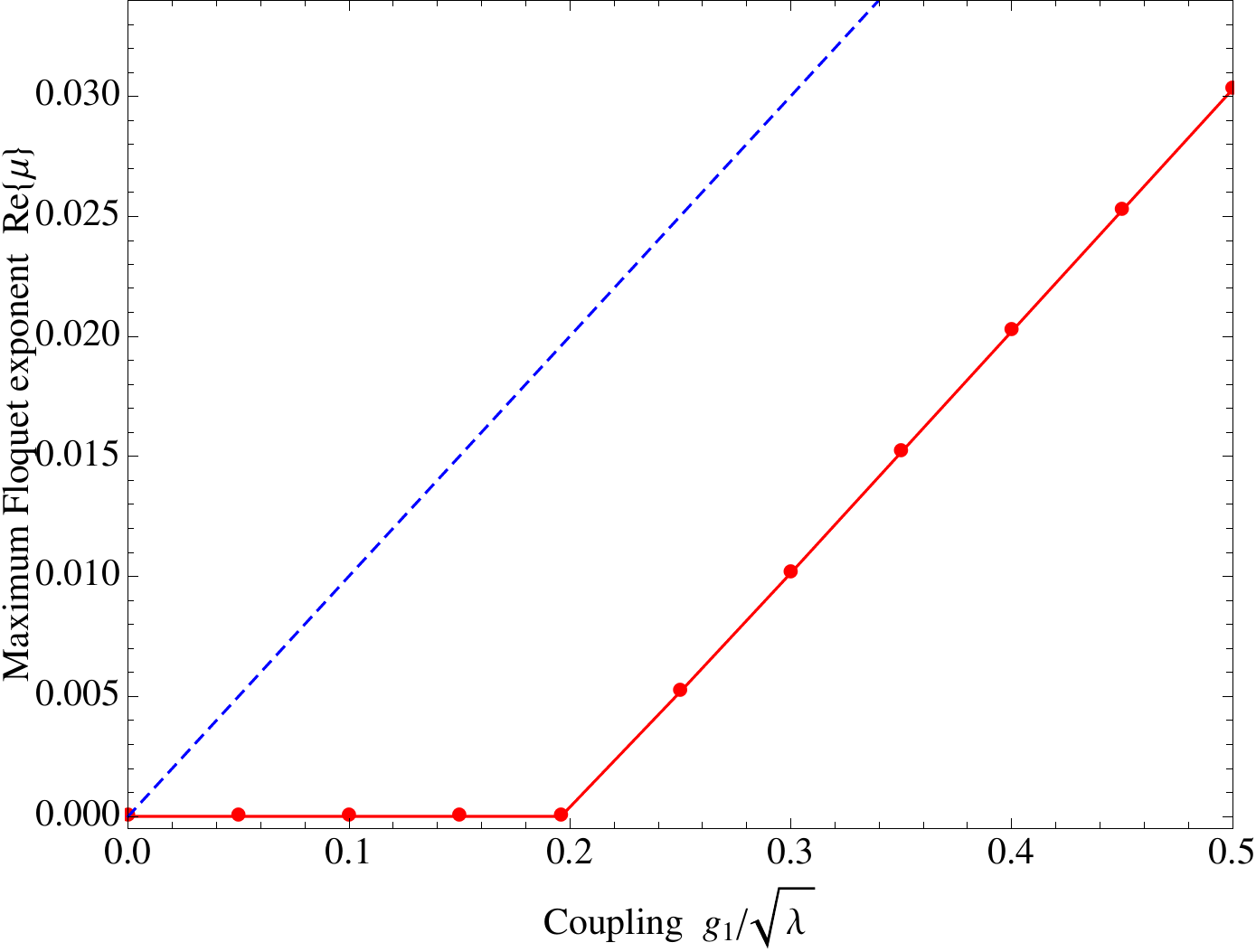}\\
\vspace{4mm}
\includegraphics[width=\columnwidth]{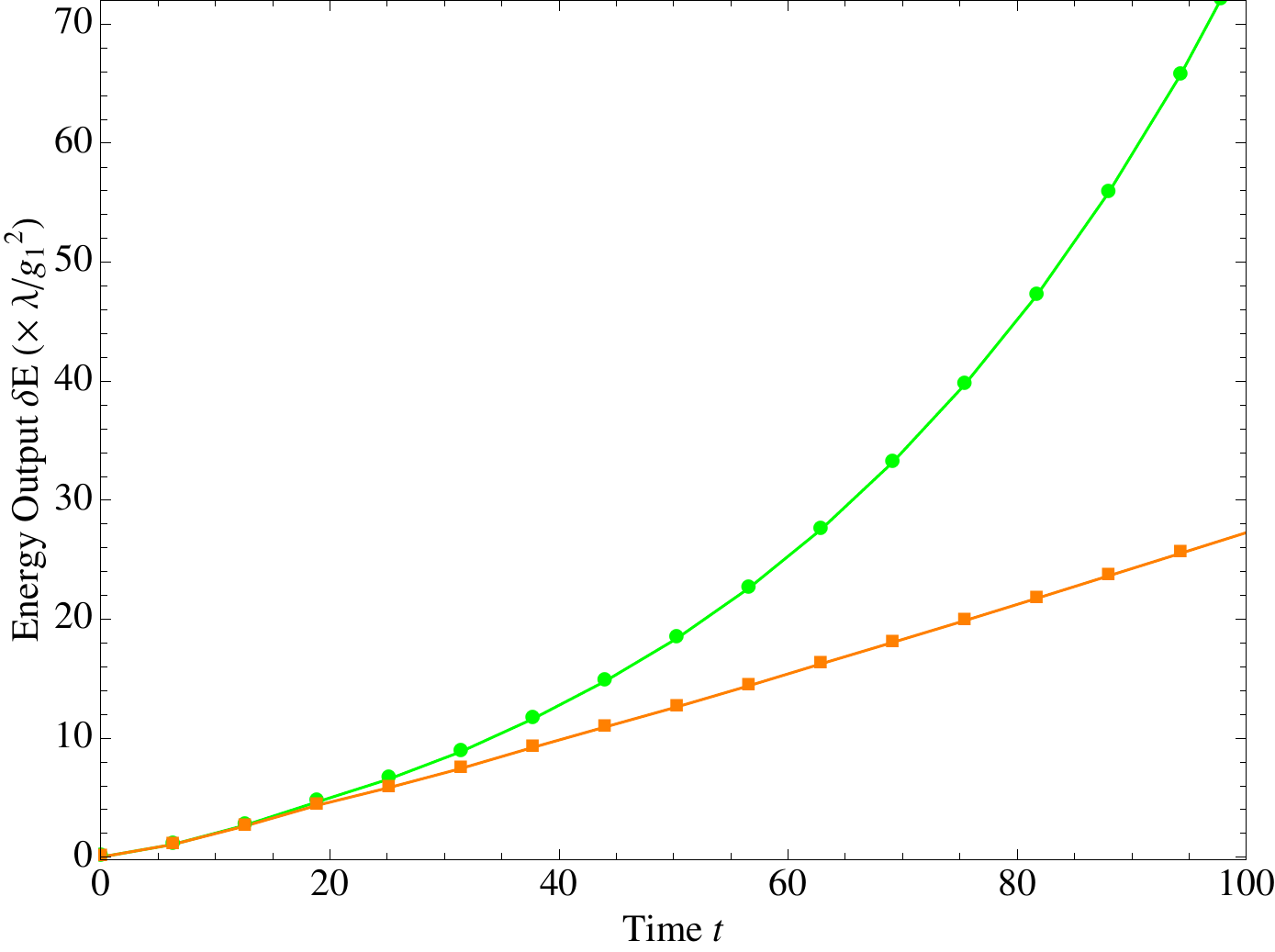}}
\caption{Top panel: the maximum value of the real part of the Floquet exponent $\mu_{\mbox{\tiny{max}}}$ as a function of coupling $g_1/\sqrt{\lambda}$ for $\epsilon=0.05$.
Lower (red) curve is for an oscillon pump $\mu_{\mbox{\tiny{max}}}$ and higher (blue-dashed) curve is for a homogeneous pump $\mu_{\mbox{\tiny{h,max}}}$.
Lower panel: total energy output $\delta E \,(\times \lambda/g_1^2)$ as a function of time for $g_1/\sqrt{\lambda}=0.1$ (orange) and $g_1/\sqrt{\lambda}=0.3$ (green).}
\label{quantumgrowth}
\end{figure}
As the figure reveals, there is a critical value of the coupling $g_1^*\approx 0.2 \sqrt{\lambda}$ governing the existence of exponential growth. For $g_1>g_1^*$ exponential growth occurs, but for $g_1<g_1^*$ it does not; in the latter regime all Floquet exponents are imaginary. This is reflected in the evolution of the energy $\delta E(t)$, which we have plotted in Fig.~\ref{quantumgrowth} (lower panel).
The lower (orange) curve is for $g_1=g_1^*/2$ and the upper (green) curve is for $g_1=3g_1^*/2$.
Hence for sufficiently small couplings the perturbative analysis is correct -- the growth is indeed linear at late times,
but for moderate to large couplings the perturbative analysis breaks down -- the growth is exponential at late times.

How can we understand this behavior? The answer resides in examining the structure of the corresponding eigenvector.
For $g_1>g_1^*$, we have numerically computed the eigenvector $\{w_k,\,z_k\}_{\mbox{\tiny{max}}}$
corresponding to the maximum Floquet exponent.
It is useful to represent this vector in position space, where it is some wave-packet. The real part of the (unnormalized) 
eigenvector is plotted in blue in Fig.~\ref{DangerousResonance} (top panel). In red we have also indicated the shape of the oscillon.
We find that the shape of the wave-packet which carries $\mu_{\mbox{\tiny{max}}}$ is approximately described by the function
\beq
\chi_{\mbox{\tiny{max}}}(x)\sim\phi_\epsilon(x)\cos(k_{\mbox{\tiny{rad}}}\,x)
\label{WP}\eeq
where $k_{\mbox{\tiny{rad}}}$ is the wavenumber we identified in the perturbative analysis ($\omega_k^2=k^2+m_\chi^2=\omega/2$). Notice that the shape of this is independent of the coupling $g_1$.
Such a result cannot make sense at arbitrarily small values of $g_1$. At sufficiently small $g_1$ {\em all} eigenvectors of the Floquet matrix $M(T)$ should be small deformations of plane waves, since we are then almost solving the Klein-Gordon equation. In particular, this means there should not be any localized wave-packet eigenvectors of the Floquet matrix. If the eigenvectors are spatially delocalized, they cannot grow exponentially, since there is nothing available to pump the wave at large distances from the oscillon.
This explains why all $\mu$ are imaginary at sufficiently small $g_1$.   
Conversely, at sufficiently large $g_1$, some solutions can exist  that are $\mathcal{O}(1)$ deviations from plane-waves, namely
the wave-packet of Fig.~\ref{DangerousResonance}. Clearly then it is inapplicable to treat this as a small perturbation from a plane wave. This explains why exponential growth can occur at sufficiently large coupling.

With this understanding, let's postdict the critical value of the coupling in this model.
If we ignore the spatial structure and treat the background oscillon as homogeneous with amplitude $\phi_\epsilon(0)$,
then $\chi$'s mode functions satisfy a Mathieu equation, whose properties are well known (let's call their Floquet exponents $\mu_{\mbox{\tiny{h}}}$). In the regime of narrow resonance,
the first instability band (connected to $\phi\to\chi+\chi$ decay) has a maximum growth rate
$\mu_{\mbox{\tiny{h,max}}}\approx g_1\phi_\epsilon(0)/2$ (plotted as the blue-dashed curve in Fig.~\ref{quantumgrowth}). On the other hand, the spatial structure of the oscillon means
that modes that are produced in the core of the oscillon will try to ``escape" at a rate set by the inverse of the oscillon's width. Let's
define an escape rate as $\mu_{\mbox{\tiny{esp}}}=1/(2R_{\mbox{\tiny{osc}}})$, where $R_{\mbox{\tiny{osc}}}$ is the oscillon's radius.
The critical $g_1$ can be estimated by the condition 
\beq
\mu_{\mbox{\tiny{h,max}}}^*\approx\mu_{\mbox{\tiny{esc}}}.
\eeq
To achieve exponential growth, we require $\mu_{\mbox{\tiny{h,max}}}\gtrsim\mu_{\mbox{\tiny{esc}}}$ in order for there to be sufficient time for growth to occur in the core of the oscillon before escaping, allowing Bose-Einstein statistics to be effective.
Using $\phi_\epsilon(0)=4\,\epsilon/\sqrt{\lambda}$ and 
$1/R_{\mbox{\tiny{osc}}}\approx\epsilon$, gives $g_1^*\approx\sqrt{\lambda}/4$,
in good agreement with the full numerical result.
This reasoning can be extended to other scenarios. For $g_1=0$, we can focus on annihilation driven by ${1\over2}g_2\phi^2\chi^2$.
In this case, study of the Mathieu equation reveals $\mu_{\mbox{\tiny{h,max}}}\approx g_2\phi_\epsilon(0)^2/8$, leading to $g_2^*\approx\lambda/(4\,\epsilon)$. Since $\lambda$ and $g_{1,2}$ are independent parameters, the regime $g_{1,2}>g_{1,2}^*$ is allowed (and easily satisfied for $\lambda\,\hbar\ll 1$, as required for massive oscillons).

\begin{figure}[t]
\center{\includegraphics[width=\columnwidth]{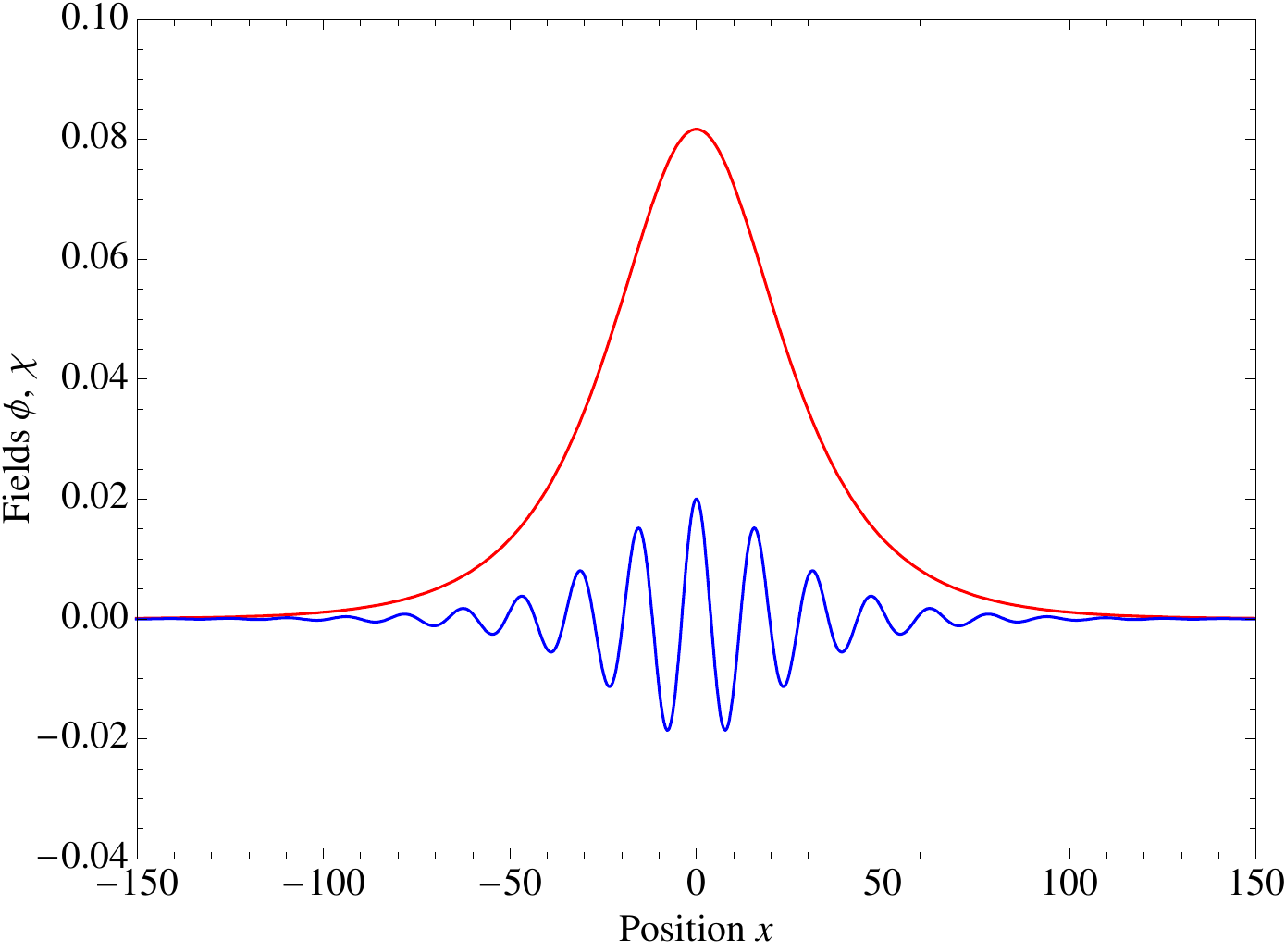}\\ \vspace{5mm}
             \includegraphics[width=\columnwidth]{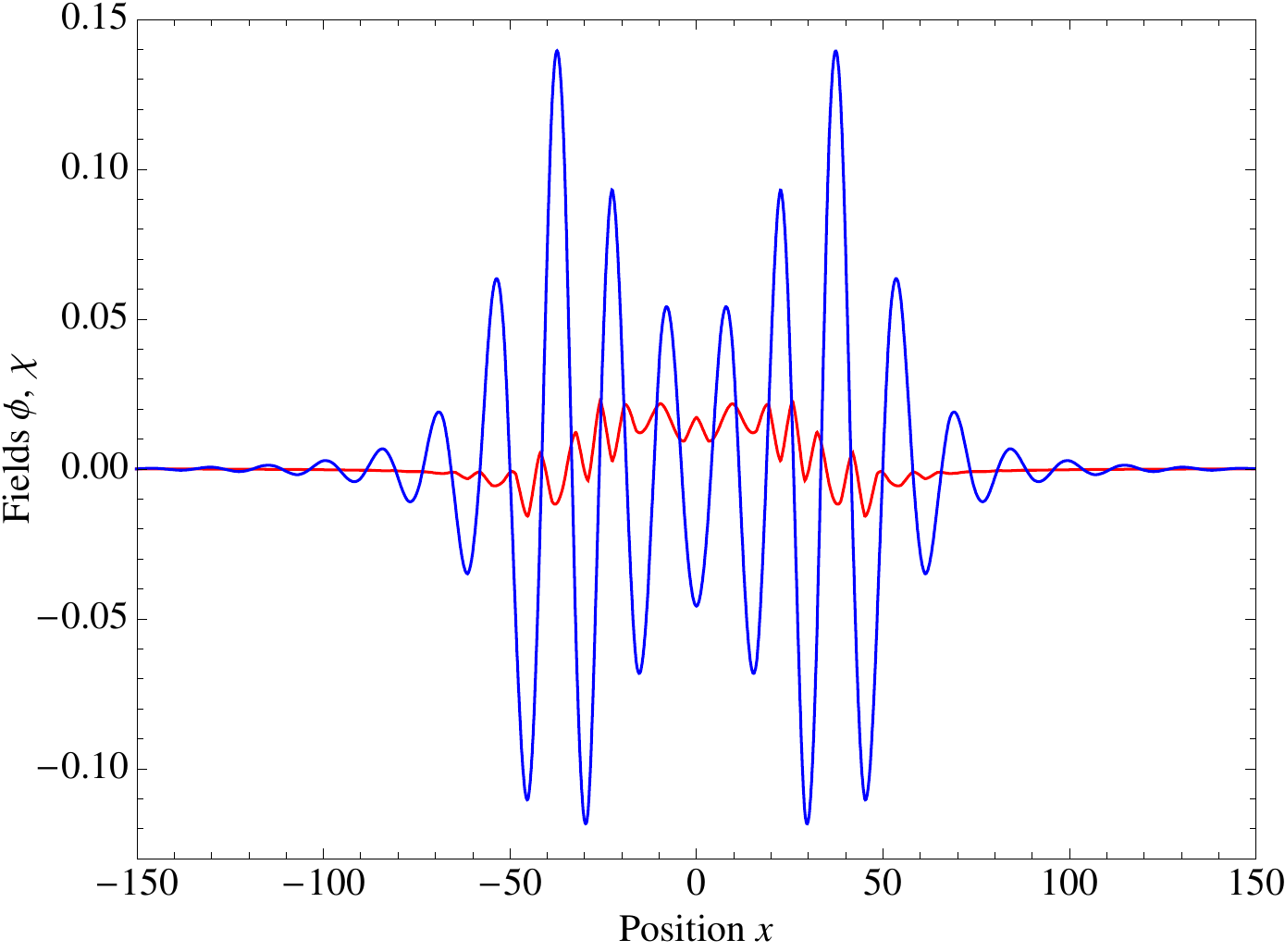}}
\caption{Top panel: The $\chi$ wave-packet which exhibits exponential growth for $g_1>g_1^*$ (blue) and the $\phi$-oscillon (red). 
Lower panel: The fields at $t=80$ after classical evolution, with $g_1=0.8\sqrt{\lambda}$, $\epsilon=0.05$, and $m_\chi=0.3$.}
\label{DangerousResonance}
\end{figure}

If the parameters are in the regime of exponential growth
it is interesting to note that substantial parametric resonance can occur from an inhomogeneous clump of energy established by oscillons. This is a form of parametric resonance -- explosive energy transfer from a localized clump to a daughter field. Of course this cannot continue indefinitely, since the oscillon has only a finite amount of energy to transfer.
Using the initial conditions of Fig.~\ref{DangerousResonance} (top panel) we have evolved the full coupled $\{\phi,\,\chi\}$ system under the classical equations of motion. We find that exponential growth in $\chi$ occurs initially and eventually this results in the destruction of the $\phi$-oscillon as seen in Figure \ref{DangerousResonance} (lower panel).

Finally, we return to the single field oscillon. In the $\lambda_5\,\phi^5$ model we originally discussed,
the resulting (generalized) Mathieu equation reveals $\mu_{\mbox{\tiny{h,max}}}\sim\lambda_5\,\phi_\epsilon(0)^3$,
leading to $\lambda_5^*\sim\lambda^{3/2}/\epsilon^2$. However, in our expansion we implicitly assumed $\lambda_5$ was $\mathcal{O}(1)$ w.r.t $\epsilon$ and therefore we should never enter the regime $\lambda_5>\lambda_5^*$.
Similarly, consider the classic $-\lambda\,\phi^4$ model. The resulting (generalized) Mathieu equation reveals $\mu_{\mbox{\tiny{h,max}}}\sim\lambda^2\phi_\epsilon(0)^4\sim\epsilon^4$.
Comparing this to $\mu_{\mbox{\tiny{esc}}}\sim\epsilon$ we see that it is impossible to obtain exponential growth for small $\epsilon$.
(This can change for the wide flat-top oscillons of Ref.~\cite{Mustafa}.) 
As far as we are aware, this is the first {\em explanation} of the stability of small amplitude oscillons 
against exponential growth of short wavelength perturbations. This fact was previously only seen empirically.
It is quite interesting that at sufficiently small amplitude, or couplings, the oscillon is stable against exponential growth in perturbations
and yet it still has modes that grow linearly with time. This occurs in the limit of degenerate eigenvalues of $M(T)$.
These modes seem relatively rare and harmless classically, but they must be integrated over in the quantum theory,
resulting in steady decay.

\section{Collapse Instabilities}
\label{LeadingQuantum}

Some oscillons are unstable to perturbations with wavelengths comparable to the size of the oscillon ($k=\mathcal{O}(\epsilon)$). 
Although this is not the focus of our paper, we would like to briefly discuss this phenomenon for completeness.
These instabilities are so prominent that they often appear in classical simulations starting from initial conditions away from the ``perfect oscillon profile", given by the expansion eq.~(\ref{phi1}). 

Using the numerical method of the previous Section we have found that for $k=\mathcal{O}(\epsilon)$ there are exponentially growing modes in $d=3$, but not in $d=1$, for the $-\lambda\,\phi^4$ theory.
Since the corresponding wavelengths are comparable to the size of the oscillon,
these instabilities are easily seen in numerical simulations starting from arbitrary initial conditions; an example is displayed in Figure \ref{TimeEvolution} for the $V_I\sim-\lambda\,\phi^4$ potential in $d=3$. In the simulation, we find that the field is growing in the core of the oscillon and we also find that it is spatially collapsing. The existence of this instability is well known in the literature.

\begin{figure}[t]
\center{\includegraphics[width=\columnwidth]{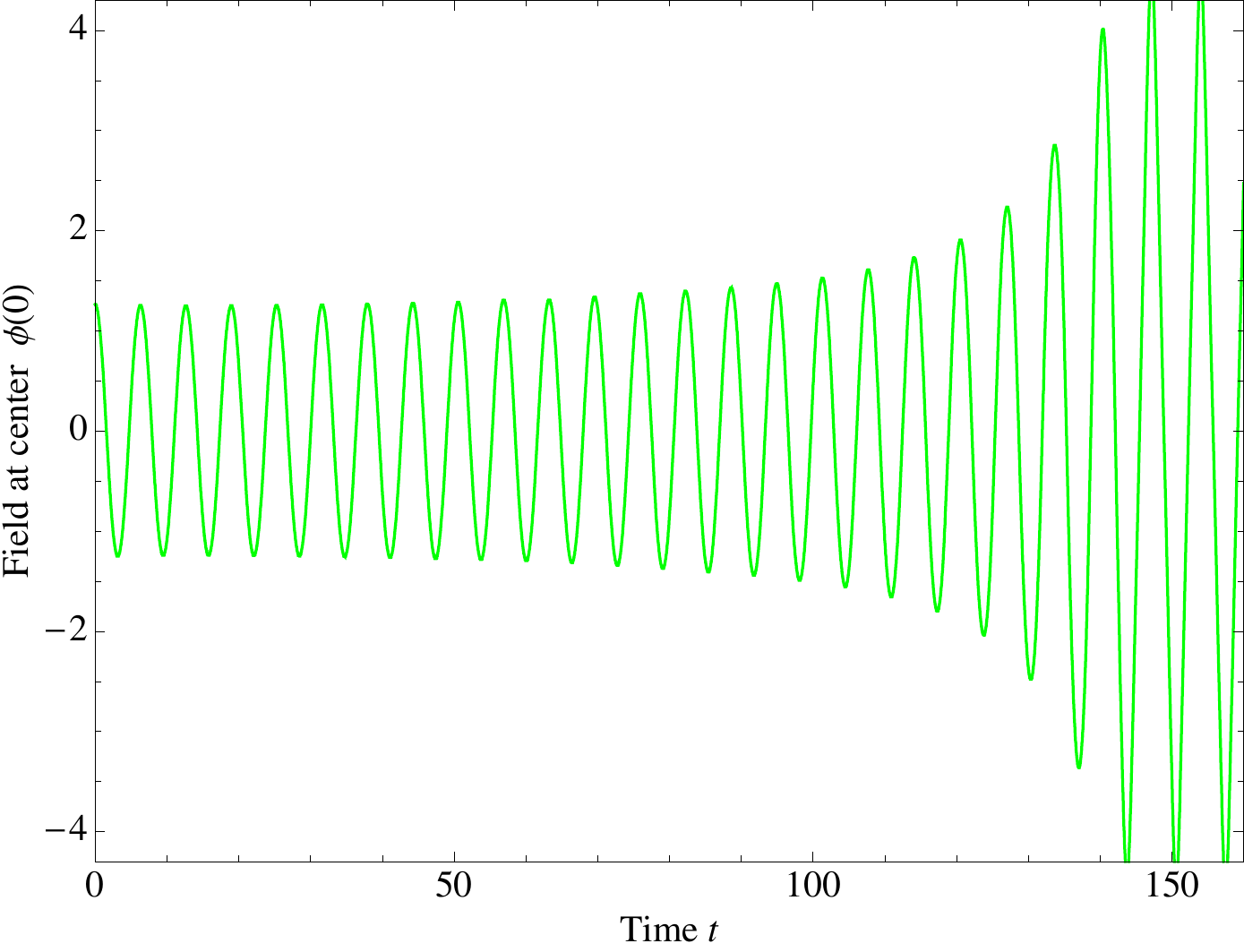}}
\caption{Field at center of oscillon over time for $V_I=-\phi^4/4!$ in $d=3$.
We choose $\epsilon=0.05$ and set up initial conditions with $\phi(0,\rho)=1.03\,\epsilon\,\phi_1(0,\rho)$, i.e., an initial profile 3\% higher than the preferred oscillon profile; this clearly causes an instability by $t\sim 150$.}
\label{TimeEvolution}
\end{figure}


To feel more confident that this is the correct behavior of the quantum theory, 
let us turn now to address this problem in a different approximation.
At some level we should consider the oscillon as a collection of $\phi$-particles in the full quantum theory \cite{Dashen:1975hd}.
Since these ``collapse" or ``self-focussing" type of instabilities occur at small wavenumbers $k=\mathcal{O}(\epsilon)$, perhaps we can interpret this as resulting from the interaction of non-relativistic particles.
The leading interaction is $2\,\phi\to2\,\phi$ scattering. 
Lets consider $V_I=\frac{1}{3!}\lambda_3\,\phi^3+\frac{1}{4!}\lambda_4\,\phi^4+\ldots$. 
At tree-level this scattering process occurs due to 4 diagrams: s, t, and u-channels generated by 2 insertions of the $\phi^3$ vertex and a single contact term from the $\phi^4$ vertex.
The matrix element is easily computed in the non-relativistic limit
$i\mathcal{M}(2\,\phi\to2\,\phi)=-i\left(\frac{5}{3}\lambda_3^2-\lambda_4\right).$
This can be recast in position space as a two particle potential $V({\bf r}_1,{\bf r_2})$ by taking the inverse Fourier transform
and multiplying by $\hbar^2/4$ due to our normalization convention. 
This gives
\bea
V({\bf r}_1,{\bf r}_2) = -{\hbar^2\over4}\left(\frac{5}{3}\lambda_3^2-\lambda_4\right)\delta^d({\bf r}_1-{\bf r}_2). 
\eea
(The s, t, and u-channels actually produce a type of Yukawa potential if the full relativistic $\mathcal{M}$ is used,
but this shouldn't modify our conclusions.)
This is a short range force and is attractive if and only if $\frac{5}{3}\lambda_3^2-\lambda_4>0$, which is a condition for the existence of oscillons. (This condition also emerges in the classical small $\epsilon$ expansion).
In the non-relativistic regime, the wavefunction for $N_\phi$ particles making up an oscillon 
will be governed by the Schr\"odinger equation 
\beq
\left(-{\hbar^2\over2m_\phi}\sum_{i=1}^{N_\phi}\nabla^2_i+\sum_{i<j}V({\bf r}_i,{\bf r}_j)\right)\psi_{N_\phi}=E_{N_\phi}\,\psi_{N_\phi}
\eeq
with the above potential $V$.  
It is known that a 
$\delta$-function potential permits unbounded solutions for $d\geq 2$ ($d=2$ is marginal, being only a logarithmic divergence, but $d\ge 3$ is a power law). 
A localized gas of $\phi$-particles, such as the oscillon, would be expected to be unstable to radial collapse under these conditions.
Hence, we expect $d> 2$ to be unstable (with $d=2$ marginal). For $d=3$ a collapse time of $\sim1/(\lambda\,\hbar\,\epsilon^2)$ is naively expected.

There is evidence in the literature \cite{Fodor:2008es,Mustafa} that oscillon stability is controlled by the derivative of the oscillon's mass $M_{\mbox{\tiny{osc}}}$ w.r.t amplitude $\phi_a$. If the derivative is positive (negative), then the oscillon appears to be stable (unstable) to collapse.
Since the canonical small amplitude oscillon satisfies $M_{\mbox{\tiny{osc}}}\sim\phi^{2-d}_a$, we obtain a consistent result.
However, this leading order behavior must break down at some point for sufficiently large amplitudes and can be affected by the inclusion of higher order terms in the potential. In fact it is known that beyond a critical amplitude no collapse instability exists in 3-d
\cite{Fodor:2008es,Mustafa} (also see the Q-ball literature \cite{Anderson:1971pt,Lee:1991ax,Belova:1997bq}). 
Collapse in 3-d is also absent in other field theories, such as the SU(2) sector of the standard model \cite{Farhi:2005rz,Graham:2006vy} and any model where $\phi_\epsilon\sim\epsilon^2$ instead of the canonical $\phi_\epsilon\sim\epsilon$ (e.g., see \cite{Fodor:2009kg}). 

In summary, the collapse of an oscillon is highly model dependent and can be avoided by operating in the appropriate number of dimensions, parameter space, or field theory. However, the radiation we computed in the previous sections is unavoidable.


\section{Conclusions}\label{Conclusions}

We have found that even though an oscillon can have a mass that is much greater than the mass of the individual quanta, the classical decay can be very different to the quantum decay
 (this point does not appear to have been appreciated in the literature, for instance see the concluding sections of Refs.~\cite{Graham:2006vy,Graham:2007ds}). 
The radiation of both classical and quantum oscillons can be understood in terms of 
forced oscillator equations, see eqs.~(\ref{classFO},\,\ref{vq1}).
We derived the frequency and wavenumber of the outgoing radiation, which were both $\mathcal{O}(1)$ in natural units. Since a classical oscillon has a spread which is $\mathcal{O}(1/\epsilon)$ in position space, it has a spread which is $\mathcal{O}(\epsilon)$ in $k$-space. Its Fourier modes are therefore exponentially small at the radiating wavenumber and hence such radiation is exponentially suppressed.
In the quantum theory, there simply {\em cannot} be modes whose amplitudes are exponentially suppressed. Instead, zero-point fluctuations ensure that all modes have at least $\mathcal{O}(\hbar)$ amplitude-squared due to the uncertainty principle. 

We derived a formula for the quantum lifetime of an oscillon $\sim 1/(\lambda\,\hbar\,\epsilon^p)$. The power $p$ is model dependent: $p=4$ in the $\phi^3+\ldots$ theory (or $-\phi^4+\phi^5+\ldots$), and $p=6$ in the $-\phi^4+\phi^6+\ldots$ theory. 
Through a Floquet analysis, we explained why the growth of perturbations of small amplitude oscillons is linear in time, 
as opposed to exponential.
The dimensionless $\lambda\,\hbar$ controls the magnitude of the decay rate, as it should for a leading order in $\hbar$ analysis. For example, the Standard Model Higgs potential has $\lambda\,\hbar\sim (m_{\mbox{\tiny{H}}}/v_{\mbox{\tiny{EW}}})^2\sim 0.1\,(m_{\mbox{\tiny{H}}}/100\,\mbox{GeV})^2$ and so this is not very small. On the other hand, the effective $\hbar$ of the QCD axion potential is 
$\lambda\,\hbar\sim(\Lambda_{\mbox{\tiny{QCD}}}/f_a)^4\sim 10^{-48}\,(10^{10}\,\mbox{GeV}/f_a)^4$ and so oscillons formed from axions, called ``axitons" in \cite{Kolb}, are governed by classical decay (ignoring coupling to other fields). 

We further considered the fate of an oscillon that is coupled to a second scalar $\chi$
and found it to either decay or annihilate with a growth in $\chi$ that can be exponentially fast, depending on parameters.
Since oscillons may form substantially in the early universe \cite{Farhi:2007wj,Copeland:1995fq} this may give rise to interesting phenomenology. At the very least, it presents a plausible cosmological scenario in which a parametric pump field exists that is qualitatively different to the homogeneous oscillations of the inflaton during p/reheating. This is a form of parametric resonance: explosive transfer of energy from a localized clump into bosonic daughter fields. We expect decay into fermions to be quite different (for discussion in the context of Q-balls, see \cite{Cohen:1986ct}). This may have some cosmological relevance. 


It appears that if a field has a perturbative decay channel, then the oscillon will eventually decay through it.
This is important because we expect most fields in nature to be perturbatively unstable, including the inflaton, p/reheating fields, Higgs, and most fields beyond the standard model. A good exception is dark matter. 
This conclusion may seem surprising given that the oscillon is a bound state of particles with a finite binding energy \cite{Dashen:1975hd}.
However, oscillons are formed from fields whose particle number is not conserved.
One could imagine a situation in which $m_\phi$ is only slightly greater than $2\,m_\chi$, and in this case the oscillon's binding energy may prevent direct decays into $\chi$'s, but this requires fine tuning and will not forbid $2\,\phi\to2\,\chi$ or $3\,\phi\to2\,\phi$ or $4\,\phi\to2\,\phi$ annihilations.


We conclude that in many scenarios an individual oscillon's lifetime will be shorter than the age of the universe at the time of production (this may prevent individual oscillons from having cosmological significance in such cases). 
Exceptions include the grand unified theory era \cite{Copeland:1995fq}, inflation \cite{Gleiser:2006te}, and axitons produced at the QCD phase transition \cite{Kolb}.
An interesting question for further study is whether oscillons can form and then decay, and then form again repeatedly, like subcritical bubbles in hot water. It is not implausible that such a process could continue over long time scales for cosmic temperatures of order the field's mass; similar to the production and disappearance of unstable particles in a relativistic plasma. This may modify cosmological thermalization.


\acknowledgments 
We would especially like to thank Mustafa Amin for helpful discussions. We also thank Andrea De Simone, Eddie Farhi, David Gosset, Saso Grozdanov, Alan Guth, Nabil Iqbal, Roman Jackiw, Aneesh Manohar, Mark Mezei, Surjeet Rajendran, Ruben Rosales, Evangelos Sfakianakis, Dave Shirokoff, Max Tegmark, and Frank Wilczek.
This work was supported by the Department of Energy (D.O.E.) under cooperative research agreement DE-FC02-94ER40818 and NSF grant AST-0134999.

  \end{document}